\documentclass[12pt,twoside, leqno]{article}

\usepackage{ifthen}

\newboolean{tab}
\setboolean{tab}{false}

\usepackage[latin1]{inputenc}
\usepackage{graphicx}
\usepackage{latexsym}
\usepackage{amssymb}
\usepackage{epsfig}
\usepackage{enumitem}
\usepackage[authoryear]{natbib}
\def\eps{\varepsilon}
\def \R{\mathbb{R}}
\def \E{\mathbb{E}}
\def \N{\mathbb{N}}

\def \T{\mathbb{T}}
\def \Cov{\mbox{Cov}}

\newcommand{\F}{\mathcal{F}}

\newcommand{\bean}{\begin{eqnarray*}}
\newcommand{\eean}{\end{eqnarray*}}
\newcommand{\bea}{\begin{eqnarray}}
\newcommand{\eea}{\end{eqnarray}}
\newcommand{\be}{\begin{eqnarray}}
\newcommand{\ee}{\end{eqnarray}}
\newcommand{\beq}{\begin{equation}}
\newcommand{\eeq}{\end{equation}}

\newcommand{\Ze}{\mathcal{Z}}

\newcommand{\Y}{\mathcal{Y}}
\newcommand{\X}{\mathcal{X}}
\newcommand{\G}{\mathcal{G}}
\newcommand{\D}{\mathcal{D}}

\newcommand{\M}{\mathcal{M}}

\newcommand{\XX}{\mathbf{X}}
\newcommand{\YY}{\mathbf{Y}}
\newcommand{\WW}{\mathbf{W}}
\newcommand{\KK}{\mathbf{K}}
\newcommand{\HH}{\mathbf{H}}
\newcommand{\MM}{\mathbf{M}}

\newcommand{\kb}{\mathbf{k}}
\newcommand{\mb}{\mathbf{m}}
\newcommand{\lb}{\mathbf{l}}
\newcommand{\eb}{\mathbf{1}}

\newcommand{\ef}{\mathcal{F}}

\begin{document}

\title{\bf Significance testing in  quantile regression}

\author{ Stanislav Volgushev$^{\normalsize\rm a,d}$\thanks{Supported by the Sonderforschungsbereich ``Statistical modelling of nonlinear dynamic processes" (SFB~823) of the Deutsche Forschungsgemeinschaft.} , Melanie Birke$^{\normalsize\rm b}$,  Holger Dette$^{\normalsize\rm a*}$, Natalie Neumeyer$^{\normalsize\rm c}$
\\
$^{\normalsize\rm a}$ Ruhr-Universit\" at Bochum \vspace{1mm}
\\
$^{\normalsize\rm b}$ Universit\"{a}t Bayreuth \vspace{1mm}
\\
$^{\normalsize\rm c}$ Universit\"{a}t Hamburg \vspace{1mm}
\\
$^{\normalsize\rm d}$ University of Illinois at Urbana-Champaign. 
$\;$\vspace{-7mm}\\
}
\date{}
\maketitle

%\author{
%{\small  Stanislav Volgushev } \\
%{\small Ruhr-Universit\"at Bochum }\\
%{\small Fakult\"at f\"ur Mathematik} \\
%{\small 44780 Bochum, Germany }\\
%{\small e-mail: stanislav.volgushev@rub.de}\\
%\and
%{\small  Melanie Birke } \\
%{\small Universit\"{a}t Bayreuth  }\\
%{\small Fakult\"{a}t f\"{u}r Mathematik, Physik und Informatik } \\
%{\small 95440 Bayreuth, Germany }\\
%{\small e-mail: melanie.birke@uni-bayreuth.de}\\
%\and
%{\small Holger Dette} \\
%{\small Ruhr-Universit\"at Bochum }\\
%{\small Fakult\"at f\"ur Mathematik} \\
%{\small 44780 Bochum, Germany }\\
%{\small e-mail: holger.dette@rub.de}\\
%\and
%{\small Natalie Neumeyer}\\
%{\small  Universit\"{a}t Hamburg}\\
%{\small   Fachbereich Mathematik} \\
%{\small   20146 Hamburg, Germany} \\
%{\small e-mail:   neumeyer@math.uni-hamburg.de}\\
%}

%\maketitle

\newtheorem{theo}{Theorem}[section]
\newtheorem{lemma}[theo]{Lemma}
\newtheorem{cor}[theo]{Corollary}
\newtheorem{rem}[theo]{Remark}
\newtheorem{prop}[theo]{Proposition}
\newtheorem{defin}[theo]{Definition}
\newtheorem{example}[theo]{Example}

\begin{abstract}
We consider the problem of testing significance of predictors in multivariate nonparametric quantile regression.
A stochastic process is proposed, which is based on a comparison of the responses with a nonparametric
quantile regression estimate under the null hypothesis. It is demonstrated that under the null hypothesis this process converges weakly to a centered Gaussian process  and the asymptotic properties of the test under fixed and local alternatives are also discussed. In particular  we show, that - in contrast to the nonparametric approach based on estimation of $L^2$-distances - the new test is able to detect local alternatives which converge to the null hypothesis with any rate $a_n \to 0$ such that $a_n \sqrt{n} \to \infty$ (here $n$ denotes the sample size). We also present a small simulation study illustrating the finite sample properties of a bootstrap version of the the corresponding Kolmogorov-Smirnov test.
\end{abstract}



\section{Introduction}
\def\theequation{1.\arabic{equation}}
\setcounter{equation}{0}

Nonparametric regression methods have become very popular in the last decades because of the fact that employing a mis-specified parametric model will typically result in inconsistent estimates and as a consequence invalid statistical inference. In recent years many
authors have developed nonparametric quantile regression estimates, which provide an attractive  supplement to least squares methods by focussing on the estimation of the conditional quantiles instead of the mean function [see \cite{chaudhuri1991},  \cite{yujon1997}, \cite{yujon1998}, \cite{detvol2008}, \cite{chefergal2010} or \cite{bondell2010} among many others]. These references mainly discuss the case of a one dimensional predictor, but
from a theoretical point of view the methods can easily be generalized to  multivariate predictors.
On the other hand it is well
known that in practical applications such  nonparametric methods suffer from the curse of dimensionality and therefore
do not yield precise estimates of conditional quantile surfaces for realistic sample sizes.
In such cases a natural and very important question is which predictor variables  are significant.

The problem   of testing significance has found considerable interest in multivariate mean regression models. \cite{gozalo1993}
 considered conditional moment tests, while \cite{yatchew1992}
constructed a test based on semi-parametric least-squares residuals.  \cite{lavvuo1996}  suggested
a directional testing procedure for discriminating between two sets of regressors without specifying the functional form of the mean regression, and \cite{racine1997}
  proposed a test based on nonparametric estimates of the partial
derivatives of the conditional mean of the response. \cite{lavvuo2000}     used the
kernel method to develop a test for the significance of a subset of explanatory variables and \cite{delgon2001}   proposed a test which is based on functionals of a $U$-process.

Because of the well known robustness properties of the conditional quantile and the fact that conditional quantiles characterize the
entire distribution it   is of particular interest to develop methods for testing significance of predictors  in  quantile regression models.
Surprisingly, in quantile regression this problem  has found much less attention. Variable selection  in the framework of linear
quantile regression models has  been recently
considered  by \cite{zouyuan2008}, \cite{wuliu2009} and \cite{belche2011} among others.   \cite{haejeoson2012}  proposed a test for
 significance  in a multivariate quantile regression model.
The work of these authors was motivated by Granger quantile causality [\cite{granger1969}] and they
employed an idea of \cite{zheng1998}, who
proposed  to transform quantile restrictions to mean restrictions. The corresponding test is based on a $U$-statistic, which estimates the
distance measure
\begin{equation}\label{hard}
\Delta= E[(P(Y \leq q_\tau(X) |X,Z)-\tau )^2 f_Z(Z)],
\end{equation}
where $Y$ denotes the response, $(X,Z)$ is the predictor, $f_Z$ the density of $Z$ and $q_\tau(X)$ the conditional $\tau$-quantile
of $Y$ given $X$. Note that the quantity $\Delta$ vanishes if and only if the conditional quantile of $Y$ given $X$ and $Z$ does not depend on $Z$.
A major drawback of this approach lies in the fact that non-parametric smoothing over both $X$ and $Z$ is needed for the construction of the estimate. This implies that the test is of very limited use when the dimension of $(X,Z)$ is larger than $3$.   Moreover, this test can only detect local alternatives converging to the null hypothesis $H_0: \Delta=0$ at a rate $n^{-1/2}h^{-(d+q)/4}$, where $d$ and $q$ are the dimensions of the predictors $X$ and $Z$, respectively, and $h$ denotes a bandwidth converging to $0$ with increasing sample size $n$.

The present paper is devoted to the problem of constructing a test for the hypothesis of the significance of the predictor $Z$, i.e. $\Delta= 0$, in the nonparametric quantile regression model, which can detect local alternatives converging to the null hypothesis at a parametric rate and at the same time does not depend on the dimension of the predictor $Z$, such that smoothing with respect to the covariate $Z$ can be avoided.
To be precise, the test proposed in this paper can detect alternatives converging to $H_0$ at any rate $a_n \to 0$ such that $a_n \sqrt{n} \to \infty$, where $n$ denotes the sample size. Our approach is based on an
empirical process, which estimates the functional
\begin{eqnarray}\label{funct}
T(x,z) &=& E[(P(Y  \leq q_\tau(X) |X,Z)) -\tau) I\{X\leq x\}  I\{Z\leq z\} ] \\
&=& E[( I\{ Y  \leq q_\tau(X)\}   -\tau) I\{X\leq x\}  I\{Z\leq z\} ] \nonumber
\end{eqnarray}
for all $(x,z)$ in the support of the distribution of the predictor $(X,Z)$, where
the inequality $X\leq x$  between the vectors $X$ and $x$ is understood as the vector of inequalities between
the corresponding coordinates and   $I\{ A \} $ denotes the
characteristic  function of the event $A$. The model, necessary notation and definition of this process
are introduced in Section \ref{sec2} and a stochastic expansion of the process $T_n(x,z)$ is established in Section \ref{sec3}.
This result allows us to obtain the weak convergence of an appropriately scaled and centered version of $T_n(x,z)$ under  the null hypothesis,  fixed and local alternatives. As a result we obtain a Kolmogorov-Smirnov or a Cramer von Mises type statistic for the hypothesis of the significance of the predictor $Z$ in the nonparametric quantile regression model.
Moreover, we are also able to extend the result to the case, where the dimension $q$ of the predictor $Z$ is growing 
with the sample size, that is $q=q_n \to \infty$ as $n \to \infty$.
 The finite sample properties of a
corresponding bootstrap test are investigated in Section \ref{sec4}. As a by-product of our theoretical analysis we also obtain new results on the uniform convergence of the conditional quantile estimator proposed by \cite{detvol2008}.
Finally all proofs, which are complicated, are deferred to an
Appendix in Section \ref{sec6}.

\section{Model, assumptions and test statistic} \label{sec2}
\def\theequation{2.\arabic{equation}}
\setcounter{equation}{0}

Let $Y$,  $X$ and $Z$ denote  one-, $d$ and $q$ dimensional random variables,  respectively, where $Y$ corresponds to the
response and  $X$ and $Z$ are the covariates. We assume  that the random variables $\{ (Y_i,X_i,Z_i)\}_{i=1,\dots,n}$ are
independent identically distributed with the same distribution as $(Y,X,Z)$.
Let $\tau\in (0,1)$ be fixed. Our aim is to test whether the predictor  $Z$ has influence on the conditional $\tau$-quantile of $Y$,
given $(X,Z)$, or whether the variable $Z$ can be omitted.  Note that this problem fundamentally differs from the question whether $Y$ is independent of $Z$ given $X$. In fact, the latter is equivalent to testing whether \textit{all} quantile curves do not depend on $Z$ as opposed to looking at a particular quantile.  Thus for fixed $\tau \in (0,1)$ we formulate the null hypothesis as
\begin{equation}\label{H0}
H_0:  E[I\{Y\leq q_\tau(X)\}-\tau\mid X,Z] = P(Y\leq q_\tau(X)\mid X,Z)- \tau = 0  \ ~a.s.,
\end{equation}
where $q_\tau (X)$ is defined as the conditional $\tau$-quantile of $Y$, given $X$, that is
\begin{equation}\label{tauq}
P(Y\leq q_\tau(X)\mid X)=\tau.
\end{equation}
It is easy to see  that  the null hypothesis (\ref{H0}) is equivalent to
$$
T(x,z) \equiv 0
$$
for all $(x,z)$ in the support of the random variable $(X,Z)$,
where the functional $T$ is defined in (\ref{funct}). This functional can be
be estimated by  the stochastic process
\begin{equation} \label{statorg}
T_n(x,z)=\frac{1}{n}\sum_{i=1}^n\big(I\{Y_i\leq \hat q_\tau(X_i)\}-\tau\big)I\{X_i\leq x\}I\{Z_i\leq z\},  ,
\end{equation}
where $(x,z) \in R_X\times R_Z$, $R_X$ and $R_Z$ denote the support of the distributions of the random variables $X$ and $Z$, respectively,
and $\hat q_\tau$ is an appropriate estimate of the conditional quantile of $Y$ given $X$, which will
be specified below. A test for the hypothesis of significance of the variable $Z$  for the $\tau$'s quantile curve of $Y$  can now easily be obtained by considering a Kolmogorov-Smirnov  or Cramer von Mises type statistic based on $T_n$ and rejecting the null hypothesis for large values of this statistic.
Throughout this paper we assume that the sets $R_X$ and $R_Z$  are compact.

In the literature, several non-parametric quantile regression estimators have been proposed [see e.g. \cite{yujon1997,yujon1998}, \cite{takleseasmo2006}, \cite{chefergal2010}  or \cite{bondell2010}  among others]. In this paper we will use an approach proposed by  \cite{detvol2008} who constructed  non-crossing estimates of quantile curves using a simultaneous inversion and isotonization of a preliminary estimator of the conditional distribution function $F_{Y|X}$ of $Y$ given $X$. For this  estimator, say $\hat F_{Y|X}(y|x;p)$, we will use
a smoothed local polynomial estimator of order $p$, see e.g. \cite{fangi1996}. Before defining this estimator, it is necessary to introduce some  notation.
\begin{itemize}
\item For $d$-dimensional vectors $x=(x(1),\ldots ,x(d)) \in \R^d $ and $\kb =(\kb(1), \ldots  \kb(d))\in \N_0^d$ define
\begin{eqnarray*}
x^\kb &:=&  (x(1)^{\kb(1)},...,x(d)^{\kb(d)}) ~,~~
\pi(x)  := x(1)\cdot x(2) \cdot ... \cdot x(d)\\
~~~\sigma(\kb)&:=&\kb(1)+...+\kb(d) ~,~~~~~~~
\kb! := \kb(1)!\cdot...\cdot \kb(d)!
\end{eqnarray*}
\item For $d$-dimensional vectors $x  \in \R^d  $, $\kb  \in \N_0^d$  and a function $K: \R\rightarrow\R$
define
\begin{eqnarray*}
\KK(x)& :=& K(x(1))\cdots  K(x(d)) ~,~~~~~~~~~~~~~~~
\KK_{h_n,\kb}(x) :=  \KK(x/h_n)\pi((x/h_n)^{\kb}) \\
\KK^{(\mb)}(x) &:= & K^{(\mb(1))}(x(1))\cdots K^{(\mb(d))}(x(d))~,~~
\KK_{h_n,\kb}^{(\mb)}(x) := \KK_{1,\kb}^{(\mb)}(x/h_n)
\end{eqnarray*}
 where $\mb =( \mb(1) \ldots ,  \mb(d)) $ is a $d$-dimensional vector with entries from $\N_0$ and $\KK^{(\ell)}$ is the $\ell$th derivative of a function $\KK$.
\item Define $N_j := \#\{\kb\in \N_0^d| \sigma( \kb ) = j\}$  as  the number of distinct $d$-tuples with size
$j$,
and denote the elements of this set by $\kb_{1,m},...,\kb_{N_m,m}$
\end{itemize}
With these notational conventions the local polynomial estimator  $\hat F_{Y|X}(y|x;p)$ of  order $p$ can be represented as [see e.g. \cite{fangi1996}]
\beq \label{def-Fdach}
\hat F_{Y|X}(y|x;p) := e_1^t(\XX^t\WW\XX)^{-1}\XX^t\WW\YY,
\eeq
where $e_1$ denotes a vector of suitable dimension with first entry one and remaining entries zero,
the matrices $\XX$, $\WW$ and the vector $\YY $ are given by
\bea \nonumber
\XX &=& \left(
\begin{array}{ccccccc}
1 & (x-X_1)^{\kb_{1,1}}& ... & (x-X_1)^{\kb_{N_1,1}} & (x-X_1)^{\kb_{1,2}} & ... & (x-X_1)^{\kb_{p,N_p}}\\
\vdots & \vdots & ... & \vdots & \vdots & ... & \vdots\\
1 & (x-X_n)^{\kb_{1,1}}& ... & (x-X_n)^{\kb_{N_1,1}} & (x-X_n)^{\kb_{1,2}} & ... & (x-X_n)^{\kb_{p,N_p}}
\end{array}
\right),
\\
\nonumber
\WW &=& \frac{1}{nh_n^d}\mbox{Diag}\Big(\KK_{h_n,0}(x-X_1),...,\KK_{h_n,0}(x-X_n)\Big),
\\
\YY &:=& \Big(\Omega \Big(\frac{y-Y_1}{d_n}\Big),...,\Omega \Big(\frac{y-Y_n}{d_n}\Big) \Big)^t,
\label{ysmooth}
\eea
and $\Omega$ denotes a smoothed version of the indicator function $I\{\cdot \leq 0\}$, that is
\beq \label{smoothkern}
\Omega (v) = \int_{-\infty}^v \omega  (u) du
\eeq
for  a given kernel $\omega$ with support $[-1,1]$.
Following \cite{detvol2008} we consider a strictly increasing distribution function $G: \mathbb{R} \to (0,1)$, a nonnegative kernel
$\kappa$ with bandwidth $b_n$, and define the functional
\beq\label{def:H}
H_{G,\kappa,\tau,b_n}(F) := \frac{1}{b_n}\int_0^1 \int_{-\infty}^\tau \kappa\Big(\frac{F(G^{-1}(u)) - v}{b_n} \Big) dvdu.
\eeq
%$$
%\widehat{H}^{-1}(\tau|x):=\frac{1}{Nb_n}\sum_{k=1}^{N}\int_{-\infty}^\tau \kappa \Bigl (\frac{\widehat{F}_{Y|X} (G^{-1}(\frac{k}{N})|x )-u}{b_n}\Bigr)du
%\: ,
%$$
%.
If $\hat F_{Y|X}$ is the estimator of the conditional distribution function defined in (\ref{def-Fdach}), it is intuitively clear that $H_{G,\kappa,\tau,b_n}(\hat F_{Y|X}(\cdot|x))$  is a consistent estimate of $H_{G,\kappa,\tau,b_n}(F_{Y|X}(\cdot|x))$.
%$$
%H_{b_n}^{-1}(\tau|x):=\frac{1}{b_n}\int_0^1\int_{-\infty}^{\tau}\kappa\Bigl(\frac{F_{Y|X}(G^{-1}(v)|x)-u}{b_n}\Bigr)dudv.
%$$
If $b_n \to 0$, this quantity %the right hand side of this equation
can be approximated as follows
\begin{eqnarray*} % \label{2.9}
H_{G,\kappa,\tau,b_n}(F_{Y|X}(\cdot|x)) &\approx&
%H^{-1}(\tau|x):=
\int_{\R}I\{F_{Y|X}(y|x)\leq \tau\}dG(y) \\
&=& \nonumber  \int_0^1I\{F_{Y|X}(G^{-1}(v)|x)\leq \tau\}dv
 \: = \: G \circ F_{Y|X}^{-1} (\tau|x),
\end{eqnarray*}
and as a consequence an estimate of the conditional quantile function $q_\tau(x)=F_{Y|X}^{-1}(\tau|x)$ can be defined by
\be  \label{qhat}
\hat q_\tau(x) :=  G^{-1}(H_{G,\kappa,\tau,b_n}(F_{Y|X}(\cdot|x))).%(\widehat{H}^{-1}(\tau|x)).
\ee

Throughout this paper, we will assume that the kernels, the function $G$ and the bandwidth parameters used to build the estimator satisfy the following conditions
\begin{enumerate}[label=(K\arabic{*})]
\item \label{as:k1} The kernel $K$ has support $[-1,1]$  and is $p+1 \geq d+2$ times continuously differentiable with uniformly bounded derivatives.
Additionally  the first $p+1$ derivatives  of $K$ vanish at the boundary points  $-1$ and $1$.
\item \label{as:k2} The function  $\omega $ in (\ref{smoothkern}) is a kernel of  order $s\geq d+1$,  has support $[-1,1]$ and  is $d$ times continuously differentiable.
Additionally  $\omega $ has uniformly bounded derivatives that vanish at the boundary points $-1$ and $1$.
\item \label{as:k3} The kernel $\kappa$ is a symmetric, positive  with support $[-1,1]$ and   has one Lipschitz-continuous derivative.
\item \label{as:g} $G: \R \rightarrow [0,1]$ is a strictly increasing distribution function such that $G,G^{-1}$ are two time continuously differentiable
\item \label{as:b1} $d_n^{2s} + h_n^{p+1} = o(1/\sqrt n)$ and $\log n/(nh_n^{3d+2}) + \log n/(nh_n^dd_n^{2d-1}) = o(1)$
\item \label{as:b2} $\frac{\log n}{nh_n^db_n^2} = o(1)$, $b_n^2 + \frac{\log n}{nh_n^db_n} + \frac{b_n \sqrt{\log n}}{\sqrt{nh_n^d}} = o(1/\sqrt n)$
\end{enumerate}

\begin{rem} \label{rem1} {\rm \cite{detvol2008} demonstrate that the choice of the distribution function $G$ has a negligible  impact  on the quality
of the resulting estimate provided that an obvious centering and standardization is performed.
Similarly, the estimate $\hat q_\tau(x)  $ is robust with respect to the choice of the bandwidth $b_n$
if it is chosen sufficiently small [see \cite{detneupil2006}].
%\textbf{Melanie: evtl. koennen wir hierzu spaeter eine kleine Simulation bringen!}
%\textbf{Additionally, we will demonstrate that the impact of the bandwidth $d_n$ is negligible as long as it is chosen sufficiently small- \textit{ich bin mir nicht so sicher ob man das machen kann}...}
}
\end{rem}

\begin{rem} \label{rem:genrear} {\rm  \cite{detvol2008} only established point-wise weak convergence of their estimator. However, for most applications such as the construction of tests on the basis of this estimator, uniform results are needed. In the present paper, we provide general inequalities for the operator $H_{G,\kappa,\tau,b_n}$ defined in (\ref{def:H}), see Lemma \ref{lem:genlin} in the Appendix. In particular, these findings allow to describe uniform properties of the quantile estimator $\hat q_\tau$ in terms of the properties of the underlying distribution function estimator $\hat F_{Y|X}$. For example, in Theorem \ref{lem1} in the appendix we exploit those bounds to derive a uniform Bahadur-type representation for the estimate $\hat q_\tau$ defined in (\ref{qhat}).
}
\end{rem}

In the following discussion it turns out to be advantageous  to consider
a generalization of the test statistic $T_n$ defined in (\ref{statorg}),
where the indicator functions $I\{X_i \leq x\}$ are replaced by indicators of
more general sets $\Theta$. To be precise  let  $\Xi$ denote   a collection of subsets of $\R^d$
 and define $\D_n := \{x \in R_X| [x-h_n \eb ,x+h_n \eb ] \subset R_X \}$ (here $ \eb $ denotes the $d$-dimensional
 vector with all entries equal to $1$), then all  theoretical developments will be based on the  statistic
\begin{equation}
\label{tn}~~~~
T_n(\Theta,z)=\frac{1}{n}\sum_{i=1}^n (I\{Y_i\leq \hat q_\tau(X_i)\}-\tau )I\{X_i\in\Theta \cap \D_n\}I\{Z_i\leq z\},\quad \Theta \in \Xi, z \in R_Z.
\end{equation}
 The intersection of the sets $\Theta \in \Xi $ with  the set $\D_n$ is needed in  the theoretical developments to exclude
  ``residuals''  $I\{ Y_i \leq \hat q_\tau (X_i) \} - \tau $ corresponding to predictors close to the boundary of $R_X$. Note that
  if $\cup_{\Theta \in \Xi} \Theta$ has a positive distance to the boundary of $R_X$, the collection of sets $\Xi_n$ will
  equal $\Xi$ whenever  $h_n$ is sufficiently small. Note also that we use the same symbol $T_n$ for the processes in (\ref{statorg}) and (\ref{tn}) but the meaning is always clear from the context. \\
Additionally to its advantages from a theoretical point of view, the consideration of a
collection  of sets that are more general than sets defined by  indicators of rectangles will for example allow
to investigate  the problem of testing the significance of the
variable $Z$  on a certain subset, say $\D \subset R_X$, that
is
\begin{equation}\label{H0D}~~~~~
H_0^{\D} :  E[ I\{Y\leq q_\tau(X)\}  I\{ X \in \D\} \mid X,Z] = P(Y\leq q_\tau(X)  \mbox{ and } X \in \D \mid X,Z) =  \tau
\end{equation}
Note that $H_0^{\D}$ means that the conditional $\tau-$quantile of $Y$ given $(X,Z)$ can be represented as a
function $q_\tau(X)$ for $X \in \D \subset R_X$. In this case  a natural choice for the collection $\Xi$
is given by  $\Xi := \{ \{X\leq t\}\cap\D|t\in \R^d\}$, but other choices are of course possible as well.

\section{Main asymptotic results}  \label{sec3}
\def\theequation{3.\arabic{equation}}
\setcounter{equation}{0}

In this section we investigate the asymptotic properties of the stochastic process defined
in (\ref{tn}). For this purpose we need some additional notation and technical
assumptions which are collected here for convenience and for later reference. \\
Define the 'error' variables as $\eps=Y-q_\tau(X)$ and $\eps_i=Y_i-q_\tau(X_i)$, $i=1,\dots,n$. We assume that
the conditional distribution function $F_{\eps|X}(\cdot|x)$  of  $\eps$ given $X=x$ has  a  density, say
$ f_{\eps|X}(y|x)$.  Note that by definition we have that $F_{\eps|X}(0|X)=P(\eps\leq 0| X)=\tau$. In particular, this identity continues to hold even if the null hypothesis is violated.
%Note that under the null hypothesis $H_0$ we have that $P(\eps\leq 0\mid X,Z)=\tau$ and hence that $F_{\eps|X}(0\mid X)=P(\eps\leq 0\mid X)=\tau$.
Throughout this paper we denote by $F_{Z|X,\eps}(z|x,e)$  the conditional distribution function of $Z$ given $(X,\eps)=(x,e)$.\\
Define $\D := \cup_{\Theta\in\Xi}\Theta$, then we assume that the data-generating process satisfies the following
conditions.

\begin{enumerate}[label=(A\arabic{*})]
%\item \label{as:b1} $d_n^{2s} + h_n^{p+1} = o(1/\sqrt n)$ and $\log n/(nh_n^{3d+2}) + \log n/(nh_n^dd_n^{2d-1}) = o(1)$
%\item \label{as:b2} $\frac{\log n}{nh_n^db_n^2} = o(1)$, $b_n^2 + \frac{\log n}{nh_n^db_n} + \frac{b_n \sqrt{\log n}}{\sqrt{nh_n^d}} = o(1/\sqrt n)$
%\item\label{as:f1} The set $\D$ is contained in the interior of the support of the joint density of $(X,Y)$
\item\label{as:f2} The conditional distribution function $F_{Y|X}(y|x)$ is $p+1$ times continuously differentiable with respect to $x,y$ and all partial derivatives are uniformly bounded  on $\R \times R_X$. The joint density of $(X,Y)$ is uniformly bounded on $R_X\times\R$. Moreover, $p\geq \max(s,d+1)$.
\item\label{as:f3} The density $f_X$ of the predictor $X$ is $d+1+n_f$ times continuously differentiable with uniformly bounded partial derivatives on $R_X$ and $n_f > d/2$. Moreover $\inf_{x\in R_X}  f_X(x) > 0 $.
%\item\label{as:f4}  $\inf_{x\in R_X}  f_X(x) > 0 $
%\item\label{as:f5} The joint density of $(X,Y)$ is uniformly bounded on $R_X\times\R$
\item\label{as:f6} There exist constants $a, C_1>0$ such that ~
$$ \inf_{(x,y): x \in R_X, |y-q_\tau(x)|\leq a} f_{Y|X}(y|x)\geq C_1.$$
\item\label{as:f7} The function $(z,x) \mapsto F_{Z|X,\eps}(z|x,0)$ is H\"{o}lder-continuous of order $\gamma>0$ with respect to $z$ and $x$ uniformly in $x\in \D$, i.e.
$$
|F_{Z|X,\eps}(s|x,0)- F_{Z|X,\eps}(t|\xi,0)| \leq C\|(s,x) - (t,\xi)\|_\infty^\gamma
$$
for some finite constant $C$.
\item\label{as:f8}~~ $\sup_{x\in \D,y\in\R,z \in\Ze} |f'_{\eps|X,Z}(y\mid x,z)|<\infty$.
\end{enumerate}

In conditions \ref{as:f2}-\ref{as:f7}, $R_X$ can be replaced by a set $\X \subset R_X$ provided that $\D \subset \X$.
Finally, the following assumptions on  the collection of sets $\Xi$ are required.

\begin{enumerate}[label=(S\arabic{*})]
\item\label{as:c1} The class of functions $\ef_1 = \{u \mapsto I\{u \in \Theta\}|\Theta \in \Xi\}$ %and $\ef_{2,c} = \{u \mapsto I\{ [u-c,u+c] \not\subset \Theta\}|\Theta \in \Xi\}$
satisfies $N_{[\, ]}(\ef_1,\eps,L^2(P_X))\leq C\eps^{-a}$
%, $N_{[\, ]}(\ef_{2,c},\eps,L^2(P_X))\leq C\eps^{-a}$
for any   sufficiently small $\varepsilon > 0$ and a constant $C$, where $N_{[\, ]}$ denotes the bracketing number
[see \cite{vaarwell1996}]
\item\label{as:c2} $\sup_{\Theta \in \Xi} P(X_i \in \Theta, \exists j: [X_i(j)-h_n,X_i(j)+h_n]\not\subset \Theta) = o(1)$ for $h_n \rightarrow 0$.
\end{enumerate}

\begin{rem}\label{rem4}{\rm  Conditions \ref{as:c1} and \ref{as:c2} are not strong and for example satisfied for  the collection of rectangles $\Xi = \{ \{s \leq X\leq t\}|s,t\in \R^d\}$     if  $X$ has a uniformly bounded density with compact support. For more details on bracketing numbers and their properties we refer to the monograph of \cite{vaarwell1996}.
}
\end{rem}

The following result gives a stochastic expansion of the process $T_n(\Theta,z)$ under general conditions, which is crucial for deriving the asymptotic properties of the process $T_n$. In particular, observe that this representation continues to hold under the alternative.

\begin{theo}\label{theo1} If the assumptions \ref{as:k1}-\ref{as:b2}, \ref{as:f2} - \ref{as:f8} and \ref{as:c1}, \ref{as:c2}
are satisfied, the process $T_n$ can be represented as
\begin{eqnarray} \label{asyexp}
~~&&~~
T_n(\Theta,z) = \frac{1}{n}\sum_{i=1}^n (I\{\eps_i\leq 0\}-\tau )I\{X_i\in\Theta_n\}(I\{Z_i\leq z\}-F_{Z|X,\eps}(z|X_i,0)) + o_P(n^{-1/2})
\end{eqnarray}
uniformly with respect to  $z \in R_Z, \Theta \in \Xi$.
\end{theo}

The proof of Theorem \ref{theo1} is complicated and given is given in the Appendix. As an immediate consequence, we obtain that under the null hypothesis $H_0$ the rescaled process $\sqrt{n}T_n(\Theta,z)$ converges weakly to a centered Gaussian process.

\begin{cor}\label{cor1}
If the assumptions of Theorem \ref{theo1} and the null hypothesis $H_0$ in (\ref{H0}) are satisfied, the process $\sqrt{n}T_n$ converges weakly in $\ell^\infty(\Xi\times R_Z)$ to a centered Gaussian process $\T$ with covariance kernel
\begin{eqnarray} \label{asycov}
~~~~~~&&
 k (\Theta_{1},y,\Theta_{2},z)  ={\rm Cov}(\T(\Theta_{1},y),\T(\Theta_{2},z))  = \tau(1-\tau)
\E \Bigl[I\{ X\in \Theta_1 \cap \Theta_2\} \\
&&   ~~~~~~~~~~~~\times
\E \Bigl[ \Bigl( I\{ Z \leq y \}  - F_{Z|X,\eps}(y|X,0) \Bigr) \Bigl( I\{ Z \leq z\}- F_{Z|X,\eps}(z|X,0) \Bigr) \Bigl|
X,\eps \Bigr]\Bigr]. \nonumber
%&=& \tau(1-\tau)
%\int I\{x\in  \Theta_1 \cap \Theta_2\}\Big(F_{Z|X,\eps} (y\wedge z\mid x,0)-F_{Z|X,\eps}(y\mid x,0)F_{Z|X,\eps}(z\mid x,0)\Big)f_X(x)\,dx,
\end{eqnarray}
\end{cor}

As a consequence of this result we obtain the weak convergence of functionals such as the Kolmogorov-Smirnov statistic
$$
K_n = \sup_{\Theta \in \Xi} \sup_{z\in R_z} | T_n(\Theta, z)|
$$
by an application of the continuous mapping theorem. In general the asymptotic distribution of $K_n$ depends on certain features of the data generating process and in the following section we will discuss bootstrap approximations for this distribution. However, in some special cases the situation simplifies substantially.

\begin{rem} {\rm
In the case where the pair $(X,\eps)$ and the covariate $Z$ are independent it follows
from (\ref{asycov}) that
\begin{eqnarray*}
&&  \Cov(\T(\Theta_{1},y),\T(\Theta_{2},z))  = \tau(1-\tau)
P( I\{X\in \Theta_1 \cap \Theta_2\} )  (F_Z (y\wedge z)-F_Z(y)F_Z(z)),
\end{eqnarray*}
where $F_Z$ is  the distribution function of the random variable $Z$ and $y\wedge z  $ denotes the vector
of  minima of the corresponding coordinates of $y$ and $z$.
If additionally  $X,Z$ are real-valued
and $\Xi = \{(-\infty,t]| t\in\R\}$, the  asymptotic covariance in Theorem \ref{theo1} reduces to
\begin{eqnarray*}
\Cov(\T((-\infty,t],y),\T((-\infty,s],z)) &=& \tau(1-\tau) F_X(s\wedge t)(F_Z (y\wedge z)-F_Z(y)F_Z(z)).
\end{eqnarray*}
Hence, for univariate independent covariates  $X$ and $Z$  with continuous distribution functions $F_X$ and $F_Z$,
respectively,  the Kolmogorov-Smirnov test is asymptotically distribution-free because in this case the statistic
$$\sqrt{n}\sup_{x\in R_X,z\in R_Z}|T_n(x,z)|=\sqrt{n}\sup_{s,t\in[0,1]}|T_n(F_X^{-1}(s),F_Z^{-1}(t))|$$
converges in distribution to $\sqrt{\tau(1-\tau)}\sup_{s,t\in[0,1]}|B(s,t)|$, where $B$ is the Kiefer-M\"{u}ller process on $[0,1]^2$, i.e. a centered Gaussian process with covariance kernel
\begin{eqnarray*}
\Cov(B(s_1,t_1),B(s_2,t_2)) &=& (s_1\wedge s_2) (t_1\wedge t_2-t_1t_2).
\end{eqnarray*}
}
\end{rem}

The result obtained in Theorem \ref{theo1} can also be used to derive the asymptotic properties of the test statistic under fixed alternatives. More precisely, the following result holds (note that under the null hypothesis, the centering term is zero, and thus this result is  a generalization of Corollary \ref{cor1}).

\begin{cor}\label{cor2} Under the assumptions of Theorem \ref{theo1} the process
\[
\sqrt{n}\Big( T_n(\Theta,z) - \int_{R_X \cap  \Theta_n }\int_{R_Z  }  \Bigl(  F_{Y |X,Z}(q_\tau(u )|u,v) - \tau  \Bigr)
I\{v \leq z \} dF_{X,Z}(u,v)\Big)
\]
converges weakly to the limiting process $\T$ defined in Corollary \ref{cor1}.
\end{cor}

\begin{rem} \label{remloc0} {\rm
A further consequence of Corollary \ref{cor2} is that the statistic $T_n$ converges
for all $\Theta \in \Xi$ and $ z\in R_Z$ in probability to the function
$$
\int_{R_X \cap  \Theta }\int_{R_Z  }  \Bigl(  F_{Y |X,Z}(q_\tau(u )|u,v) - \tau  \Bigr)
 \Bigl(  I\{v \leq z \} - F_{Z|X,\eps}(z|u ,0)\Bigr)
f_X(u) f_Z(v) du dv .
 $$
 Consequently, if $\Xi$ contains sufficiently many sets (for example, if $\Xi = \{ (-\infty, x]~|~x \in \R^d\} $), the test is
 consistent. In order to obtain the asymptotic distribution of the test statistic under local alternatives of the form
 \be
 \label{localt}
 F^{(n)}_{Y |X,Z}(q_\tau^{(n)}(u)|u,v) = \tau + a_n h(u,v)
 \ee
  a result on the asymptotic behavior of $T_n(\Theta,z)$  is required
 when the data are generated from triangular arrays. A closer look at the proofs in the appendix shows that such a result does indeed hold under suitable modifications of the conditions in Theorem \ref{theo1}. The details are omitted for the sake of brevity. In particular, a test based on the Kolmogorov-Smirnov test statistic will detect all local alternatives for which the quantity
\[
K_n= \sup_{\Theta_n,z} \Big| \sqrt{n}\int_{R_X \cap  \Theta_n }\int_{R_Z  }  \Bigl(  F^{(n)}_{Y |X,Z}(q_\tau^{(n)}(u)|u,v) - \tau  \Bigr)
I\{v \leq z \} dF^{(n)}_{X,Z}(u,v) \Big|
\]
diverges to infinity (the superscript is used to indicate that the corresponding quantities depend on $n$). For example $K_n \to \infty$ in probability if $\Xi = \{ (-\infty, x]~|~x \in \R^d\}$ and $F^{(n)}_{Y |X,Z}(q_\tau^{(n)}(u)|u,v) = \tau + a_n h(u,v)$ for some function $h$ that is not identically zero on $R_X\times R_Z$ and sequence $a_n$ with $a_n\sqrt{n} \rightarrow\infty$. This means that the test can detect alternatives
converging  to the null hypothesis at rates which are ``larger but arbitrarily close'' to the parametric rate $n^{-1/2}$. Moreover, the test will have an asymptotically non-trivial power against many local alternatives that converge to zero at the exact parametric rate $n^{-1/2}$.
}
\end{rem}

\begin{rem}
{\rm{
We now give a brief discussion of the properties of the proposed test statistic when alternatives of increasing dimension are considered, i.e. when the dimension of the predictor $Z$, say $q_n$, varies with $n$. Consider the additional assumption
\begin{enumerate}
\item[(Z)] The $L^2$ covering numbers of the classes of functions
$$\{x \mapsto F_{Z|X,\eps}(z|x+s,0)| z \in \Ze, \|s\|_\infty \leq a  \}$$
 and $\{\xi \mapsto I\{\xi \leq z\}|z\in\Ze\}$ are bounded by $C_1(C_2/\eps)^{k_n}$ for some finite constants $C_1,C_2$.
\end{enumerate}
Note that   assumption (Z) holds with $k_n = q_n$ if for each $n$ the predictor $Z$ given $(X,\eps)$ has a conditional density $f_{Z|X,\eps}$ that satisfies
$$\sup_z |f_{Z|X,\eps}(z|x_1,0) -f_{Z|X,\eps}(z|x_2,0)| \leq C \|x_1-x_2\|$$
 for a finite constant $C$ independent of $n$.
Under assumptions (K1)-(K6), (A1)-(A3), (Z), (A5), (S1), (S2) it is possible to prove that
\begin{eqnarray} \nonumber
~~&&~~
T_n(\Theta,z) = \frac{1}{n}\sum_{i=1}^n (I\{\eps_i\leq 0\}-\tau )I\{X_i\in\Theta_n\}(I\{Z_i\leq z\}-F_{Z|X,\eps}(z|X_i,0)) + o_P\Big(\frac{k_n}{n^{1/2}}\Big),
\end{eqnarray}
uniformly with respect to  $z \in R_Z, \Theta \in \Xi$. In particular, this result implies
\[
\sqrt{n}\Big( T_n(\Theta,z) - \int_{R_X \cap  \Theta_n }\int_{R_Z  }  \Bigl(  F_{Y |X,Z}(q_\tau(u )|u,v) - \tau  \Bigr)
I\{v \leq z \} dF_{X,Z}(u,v)\Big) = O_P\Big( {k_n}\Big).
\]
Consequently, the test is  able to detect local alternatives converging to the null hypothesis with any rate $a_n$, such that $\frac {a_n}{k_n} \sqrt{n}\to \infty$  when the sample size and dimension $k_n$ of $Z$ is increasing.
}
}
\end{rem}

\begin{rem}
\label{fehler}
{\rm  \cite{haejeoson2012} investigated  an alternative test for the hypothesis (\ref{H0}) based on ideas from \cite{fanli1996} in combination with a modification which was originally proposed by \cite{zheng1998}.
Their test is based on    the  statistic
$$
J_n = \frac {1}{n(n-1)g_n^d} \sum_{i,j, i \neq j} L ((Z_i - Z_j)/g_n )  (I \{ Y_i \leq \hat Q (\tau|X_{i})\}  -\tau )
(I \{ Y_j \leq \hat Q (\tau|X_{j})\}  -\tau )
$$
where $L$ is a kernel and $g_n$ is a bandwidth converging to $0$ with increasing sampling size. These authors claimed that a normalized version of this test statistic converges to a normal distribution. It should be pointed out here that the proof in this paper is not correct.
The basic argument of  \cite{haejeoson2012} consists in the statement that the fact
$$
\sup_x \mid \hat Q_\tau ( x) - Q_\tau ( x) \mid \leq C_n
$$
results in the estimate
\begin{equation} \label{falsch}
J_{nU} \leq J_n \leq J_{nL},
\end{equation}
where the statistics $J_{nU}$ and $J_{nL}$ are defined by
\begin{eqnarray*}
J_{nU} &=& \frac {1}{n(n-1)g^d} \sum_{i \neq j}  L ((Z_i - Z_j)/g )  \varepsilon_{iU} \varepsilon_{jU}, \\
J_{nL} &=& \frac {1}{n(n-1)g^d} \sum_{i \neq j}  L ((Z_i - Z_j)/g )  \varepsilon_{iL} \varepsilon_{jL},
\end{eqnarray*}
and $\varepsilon_{iU}= I \{ Y_i + C_n \leq Q_\tau ( X_{i})\} - \tau, \ \varepsilon_{iL}= I \{ Y_i + C_n \leq Q_\tau ( X_{i}) \} - \tau$ (see equation (A.11-3) in this paper). A simple calculation shows that this conclusion is not correct and in fact the inequality (\ref{falsch}) does not hold. It turns out that the proof of Theorem 1 in \cite{haejeoson2012} can not be corrected easily.\\
 Even if the gap in the proof would be closed, the test of \cite{haejeoson2012} still has two major drawbacks. First, it requires non-parametric smoothing with respect to the covariate $Z$. Second, it can only detect local alternatives converging to the null hypothesis at a rate $n^{-1/2}h^{-(d+q)/4}$ which is slower than the rate $b_n n^{-1/2}$ for any $b_n\rightarrow\infty$ detected by the test proposed in this paper and additionally depends on the dimension of the covariates. }
\end{rem}

\section{Bootstrap and simulation results}  \label{sec4}
\def\theequation{4.\arabic{equation}}
\setcounter{equation}{0}
%%%%%%%%%%%%%%%%%%%%%%%%%%%%%%%%%%%%%%
%%%%%%%%%%%%%%%%%%%%%%%%%%%%%%%%%%%%%%

In general the limit distribution derived in Theorem \ref{theo1} depends on certain features of the data generating process
which are difficult to estimate. For this reason
 we  discuss  in this section bootstrap methods that are suitable to mimic the distribution of test statistics
based on $T_n$ under the null hypothesis. To be precise, let   $P^*$ denote the conditional probability $P(\cdot\mid \Y_n)$, given the original sample $\Y_n=\{(Y_i,X_i,Z_i)\mid i=1,\dots,n\}$, and denote by $\E^*$ and $\Cov^*$ the corresponding conditional expectation and covariance.
Several residual wild bootstrap  approximations have been proposed in the literature for quantile regression analysis [see \cite{sun2006}  or \cite{fenghehu2011}]. However, the residual wild bootstrap does not yield  a valid approximation of the limiting distribution
 in the present context because it does not lead to an expansion of the bootstrap process analogous to the one given for $T_n$
 in Theorem \ref{theo1}. %In the following we discuss  two different bootstrap methods, which yield a valid approximation of the limiting distribution.

As alternative we consider the idea of process-based wild bootstrap as considered by
 \cite{delgon2001}  or \cite{hezhu2003}. To this end recall the definition of the ``residuals'' $\hat\eps_i=Y_i-\hat q_\tau(X_i)$, where $\hat q_\tau$ denotes an estimator for the conditional $\tau$-quantile of $Y_i$, given $X_i$, define
 $\hat \tau = \sum_{j=1}^n I\{\hat\eps_j\leq 0\}/n$ and introduce independent identically distributed  Bernoulli random variables
 $B_1,\ldots , B_n$ with success probability $\hat \tau$, which are independent of the original data. Define the bootstrap process as
%\[
%T_n^*(\Theta,z) = \frac{1}{n}\sum_{i=1}^nv_i\Big(I\{Y_i\leq \hat q_\tau(X_i)\}-\tau\Big)I\{X_i\in\Theta\}\Big(I\{Z_i\leq z\}-\hat F_{Z|X,\eps}(z|X_i,0)\Big),
%\]
\[
T_n^*(\Theta,z) = \frac{1}{n}\sum_{i=1}^n (B_i-\hat\tau)I\{X_i\in\Theta\}\Big(I\{Z_i\leq z\}-\hat F_{Z|X,\eps}(z|X_i,0)\Big),
\]
where
\begin{equation}
\label{fzx}
\hat F_{Z|X,\eps}(\cdot|x,y)=\frac{\sum_{j=1}^n I\{Z_j\leq \cdot\}L(\frac{X_j-x}{a})N(\frac{\hat\eps_j-y}{e})}{\sum_{j=1}^n L(\frac{X_j-x}{a})N(\frac{\hat\eps_j-y}{e})}
\end{equation}
denotes a kernel estimator for the conditional distribution $F_{Z|X,\eps}(\cdot|x,y)$. Here, $L$ and $N$ denote $d$- and  one-dimensional kernel functions and $a$ and $e$ corresponding bandwidths converging to $0$ with increasing sample size. For the sake of brevity we do not consider conditional weak convergence of the process $T_n^*$ in detail, but note that $E^*[T_n^*(\Theta,z) ]=0$ and under the null hypothesis $H_0$  (and under suitable regularity conditions)
the conditional covariance $n\Cov ^*(T_n^*(\Theta_1,y),T_n^*(\Theta_2,z))$ converges  in probability  to the covariance $\Cov(T(\Theta_1,y),T(\Theta_2,z))$ as defined in Theorem \ref{theo1}.\\
In our numerical investigations, it turned out that the asymptotic representation  (\ref{asyexp}) for the process
defined in (\ref{statorg})
 is not very accurate for small sample sizes. We thus considered a slightly modified version of this process, that is
\[
\tilde T_n(x,z)=\frac{1}{n}\sum_{i=1}^n\big(I\{Y_i\leq \hat q_\tau(X_i)\}-\hat\tau\big)I\{X_i\leq x\}(I\{Z_i\leq z\} - \hat F_Z(z))
\]
where $\hat F_Z(z)$ denotes the empirical distribution function of $Z_1,...,Z_n$, which provided much better results for moderate sample sizes.
As motivation for this approach, observe that under both the null hypothesis   and the alternative, we have
\[
D_x := \frac{1}{n}\sum_{i=1}^n\big(I\{Y_i\leq \hat q_\tau(X_i)\}-\hat\tau\big)I\{X_i\leq x\} = o_P(n^{-1/2}), \quad \hat\tau = \tau + o_P(n^{-1/2}) 
\]
uniformly with respect to $x$ as can be seen by taking a closer look at the proofs of the main results
in the Appendix. Thus the additional correction term 
\[
\delta_{x,z} := D_x\hat F_Z(z) + \frac{\hat\tau-\tau}{n}\sum_{i=1}^n I\{X_i\leq x\}I\{Z_i\leq z\}
\]
vanishes asymptotically (uniformly with respect to $x,z$) under both the alternative and the null hypothesis. If, on the other hand, $\delta_{x,z}$ is relatively large because the sample size is small, the correction term $\delta_{x,z}$ induces an additional centering (the factor $\hat F_Z(z)$ corresponds to the amount of non-zero indicators $I\{Z_i\leq z\}$). \\
The simulation results described below confirm that this is a sensible approach.\\
For the calculation of the test statistic
\be \label{ktilde}
\tilde K_n= \sup_x \sup_z \mid \tilde T_n (x,z) \mid
\ee
based on the process $\tilde T_n$, we use local polynomial estimators of order two [see (\ref{def-Fdach})].
The bandwidth $h_n$ of this estimator is chosen as $h_n := (\hat \sigma^2/2n)^{13/50}$ where $\hat\sigma^2$ denotes the variance estimate of \cite{rice1984} from the sample $\{ (X_i,Y_i)|~i=1,\dots , n\}$ [see \cite{yujon1997} for a related approach]. The bandwidths used in (\ref{ysmooth}) and (\ref{fzx}) are chosen as $d_n = a = e = h_n$, while the choice of $b_n$ in (\ref{def:H}) is even less critical [see also \cite{detvol2008}] and we
use $b_n=h_n^3$. In fact, in the simulations it turned out that the power and size properties of the test are rather insensitive with respect to the bandwidth choice, see table \ref{tab:bdw05} and related discussion in the next paragraph. The function $\omega$ in (\ref{smoothkern}) is chosen as $\omega(x) := (15/32)(3 - 10x^2 + 7x^4)I\{|x|\leq 1\}$, which is a kernel of order $2$ [see \cite{gamuma1985}]. The function $\kappa$ in (\ref{def:H}) is  defined as Epanechnikov kernel while all  other kernels are Gaussian kernels. For the choice of
 the distribution function $G$ in (\ref{def:H}) we follow the procedure described in \cite{detvol2008} who suggested a
  normal distribution such that the $5\%$ and $95\%$ quantiles coincide with  the corresponding empirical quantities of the sample $Y_1,...,Y_n$.
\subsection{Simulation results}
We simulate data from the location scale model
\be \label{mod}
Y_i = q_{j}(X_i,Z_i) + s_k(X_i,Z_i)\varepsilon_i,
\ee
$j,k=1,\ldots,4$ with the following quantile and scale functions
\begin{eqnarray} \label{loc}
q_{1}(x,z)&=&\exp(2x^2)~,~
q_{2}(x,z)=(x-0.5)^2\\
\nonumber
q_{3}(x,z)&=&\exp(2x^2)z^2~,~
q_{4}(x,z)=\sin(2\pi(x+z))
\end{eqnarray}
and
\begin{eqnarray}\label{scale}
s_1(x,z)&=&0.5(x+0.2)~,~
s_2(x,z)=0.5(\sin(x)+1.2)\\
\nonumber
s_3(x,z)&=&0.5(z+0.2)~,~
s_4(x,z)=0.5\sqrt{(x+0.2)(z+0.2)}.%6
\end{eqnarray}
The random variables $X$ and $Z$ are independent and uniformly distributed on the interval $[0,1]$ while $\varepsilon$ is standard normal. We consider the cases $\tau=0.5$ and $\tau=0.25$. All reported  results  are  based on $1000$ simulation runs with $300$ bootstrap replications.
\begin{table}
\begin{center}
\begin{tabular}{|c|l|ll|ll|ll|}
\hline
&&\multicolumn{2}{|c|}{$\alpha=0.025$}&\multicolumn{2}{|c|}{$\alpha=0.05$}&\multicolumn{2}{|c|}{$\alpha=0.1$}\\ \cline{3-8}
$\tau$ & $(k,l)$&$n=50$&$n=100$ &$n=50$&$n=100$ &$n=50$&$n=100$ \\\hline
&(1,1)&0.037&0.035&0.053&0.061&0.102&0.111\\
&(1,2)&0.026&0.025&0.044&0.048&0.090&0.101\\
&(1,3)&0.041&0.027&0.069&0.066&0.132&0.127\\
0.5&(1,4)&0.040&0.033&0.060&0.059&0.120&0.121\\
&(2,1)&0.036&0.031&0.068&0.057&0.122&0.106\\
&(2,2)&0.024&0.028&0.051&0.046&0.092&0.085\\
&(2,3)&0.037&0.025&0.057&0.059&0.132&0.114\\
&(2,4)&0.027&0.024&0.050&0.047&0.109&0.093\\
\hline\hline
&(1,1)&0.024&0.019&0.044&0.035&0.089&0.082\\
0.25&(1,2)&0.024&0.019&0.044&0.037&0.089&0.092\\
&(2,1)&0.027&0.025&0.047&0.052&0.102&0.105\\
&(2,2)&0.016&0.022&0.036&0.048&0.089&0.101\\ \hline
\end{tabular}
\end{center}
\caption  {\label{sizepower05}
\it Simulated rejection probabilities of  the bootstrap test (\ref{boottest}) for  significance of the variable $z$
in the quantile regression model (\ref{mod})  for $\tau=0.5$ (upper part) and $\tau=0.25$ (lower part) under
various null hypotheses. The pair $(k,l)$ corresponds to the location function $q_k$ and scale function $s_\ell$ specified in (\ref{loc}) and (\ref{scale}), respectively.}
\end{table}

\begin{table}
\begin{center}
\begin{tabular}{|c|l|ll|ll|ll|}
\hline
& &\multicolumn{2}{|c|}{$\alpha=0.025$}&\multicolumn{2}{|c|}{$\alpha=0.05$}&\multicolumn{2}{|c|}{$\alpha=0.1$}\\ \cline{3-8}
$\tau$& $(k,l)$&$n=50$&$n=100$&$n=50$&$n=100$&$n=50$&$n=100$\\
\hline
&(3,1)&0.999&1.000&1.000&1.000&1.000&1.000\\
&(3,2)&0.756&0.983&0.815&0.989&0.886&0.997\\
&(3,3)&0.997&1.000&0.999&1.000&0.999&1.000\\
0.5 &(3,4)&1.000&1.000&1.000&1.000&1.000&1.000\\
&(4,1)&0.082&0.197&0.142&0.311&0.252&0.519\\
&(4,2)&0.034&0.070  &0.067&0.119 &0.138&0.237  \\
&(4,3)&0.089&0.176  &0.134 &0.279 &0.226&0.488  \\
&(4,4)&0.070&0.203  &0.123&0.321 &0.218&0.508 \\
\hline\hline
&(1,3)&0.099&0.240&0.163&0.325&0.245&0.459\\
&(1,4)&0.044&0.078&0.086&0.133&0.155&0.225\\
&(2,3)&0.139&0.295&0.204&0.405&0.332&0.540\\
&(2,4)&0.06&0.089&0.106&0.152&0.176&0.232\\
&(3,1)&0.935&1.000&0.971&1.000&0.988&1.000\\
0.25&(3,2)&0.464&0.857&0.591&0.913&0.725&0.954\\
&(3,3)&0.792&0.990&0.873&0.996&0.934&0.999\\
&(3,4)&0.900&1.000&0.948&1.000&0.975&1.000\\
&(4,1)&0.027&0.054&0.055&0.103&0.111&0.229\\
&(4,2)&0.019&0.031&0.034&0.061&0.078&0.132\\
&(4,3)&0.022&0.051&0.043&0.091&0.104&0.176\\
&(4,4)&0.021&0.054&0.053&0.093&0.104&0.195\\ \hline
\end{tabular}
\end{center}
\caption
 {\label{size_power_025}
 \it Simulated rejection probabilities of  the  bootstrap test (\ref{boottest}) for  significance of the variable $z$
in the quantile regression model (\ref{mod}) for $\tau=0.5$ (upper part) and $\tau=0.25$ (lower part) under
various alternatives. The pair $(k,l)$ corresponds to the location function $q_k$ and scale function $s_\ell$ specified in (\ref{loc}) and (\ref{scale}), respectively.}
\end{table}

The bootstrap test (at level $\alpha$) rejects the null hypothesis that the variable $Z$ is not significant, whenever
\be \label{boottest}
\tilde K_n > K^*_{n,1-\alpha}
\ee
where $\tilde K_n$ is defined in (\ref{ktilde}) and $K^*_{n,1-\alpha}$  denotes the $(1-\alpha)$ bootstrap quantile of the Kolmogorov-Smirnov test statistic.

The rejection probabilities of this test under the null hypothesis are shown in  Table  \ref{sizepower05} for  the  $50\%$ and $25\%$ quantile. Note
that  different pairs of location and scale functions in (\ref{loc}) and (\ref{scale}) correspond to the null hypothesis for $\tau=0.5$ and $\tau=0.25$
(more precisely the models  defined by the pairs  $(1,3)$, $(1,4)$, $(2,3)$ and $(2,4)$ correspond to  the null hypothesis if $\tau=0.5$ but
to the alternative if $\tau=0.25$).
We observe from Table  \ref{sizepower05}  that the level is usually approximated very well. For  $\tau=0.25$ there exist some  cases
where the test is slightly  conservative .

The corresponding results for various alternatives are displayed in Table  \ref{size_power_025} and we observe a reasonable power for most
cases. The power for $\tau=0.25$ is always smaller  than the power for $\tau=0.5$. This corresponds to intuition because the $25\%$-quantile
 is more difficult to estimate than the median.  The power of the test is smaller for alternatives corresponding to the location function $q_{4}(x,z)=\sin(2\pi(x+z))$ if the
 sample size is $n=100$. However, if the the sample size is     larger, the test also detects the alternatives with reasonable probability. For example
 if $n=200$ and $\tau=0.5$  the simulated rejection probabilities of the bootstrap test at level $5\%$ for the alternatives  $(4,2)$,  $(4,3)$
 and  $(4,4)$ are given by $ 0.319$, $0.795$ and  $0.821$, respectively.  \\
 Next we study the impact of the choice of the bandwidth on size and power of the bootstrap test. For this
 purpose we consider the sample size $n=50$ and bandwidths $0.05$, $0.10$, $0.15$, $0.20$, $0.25$, $0.30$, $0.35$, $0.40$, $0.45$
  and $0.50$.
 The  results  for   model  $(1,2)$ and $(3,2)$ corresponding to the null hypothesis and alternative, respectively,
are  summarized in Table  \ref{tab:bdw05}.  We observe that the level and power are rather stable with respect to different choices of the bandwidth. Simulations for other scenarios yield similar results and are not shown for the sake
of brevity.

\begin{table}
\begin{center}
\begin{tabular}{|c|l|llllllllll|} \hline
$\tau$&$h$&0.05& 0.10& 0.15& 0.20& 0.25& 0.30&0.35&0.4&0.45&0.5\\\hline
0.5 &(1,2)&0.037&0.036&0.037&0.037&0.047&0.054&0.061&0.046&0.047&0.043\\
&(3,2)&0.238&0.301&0.361&0.389&0.388&0.385&0.381&0.389&0.412&0.404
\\\hline
\hline
0.25 & (1,2)&0.017&0.031&0.037&0.033&0.031&0.048&0.042&0.049&0.041&0.053\\
& (3,2)&0.113&0.160&0.210&0.210&0.237&0.250&0.262&0.246&0.262&0.260 \\ \hline
\end{tabular}
\end{center}
\caption{\label{tab:bdw05}
\it Simulated rejection probabilities of the bootstrap test (\ref{boottest}) for various bandwidths. The sample
size is  $n=50$ and the lower and upper part correspond to the  $50\%$ and $25\%$ quantile, respectively.
The pair $(k,l)$ corresponds to the location function $q_k$ and scale function $s_\ell$ specified in (\ref{loc}) 
and (\ref{scale}), respectively.}
\end{table}

\begin{table}
\begin{center}
\begin{tabular}{|l|lll|} \hline
$\alpha$ &0.025&0.050&0.100\\\hline
$q_1$&0.026&0.042&0.096\\
$q_2$&0.998&1.000&1.000 \\ \hline
\end{tabular}
\end{center}
\caption{\label{tab:2D}
\it Simulated rejection probabilities of the bootstrap test (\ref{boottest}) for the significance of a two dimensional predictor
in median regression.  The models are defined in (\ref{twodim}), the  sample size is  $n=50$  and the upper (lower) 
row corresponds to the null hypothesis
(alternative)}
\end{table}

We conclude our numerical study with a brief investigation of a two dimensional predictor, say $Z=(Z_1,Z_2)$.
Because the method proposed in this paper does not require smoothing in the $Z$-direction, the results should not be seriously
affected, if the dimension of $Z$ is larger. To be precise we consider two different location functions
\begin{eqnarray} \label{twodim}
q_1(x,z_1,z_2)&=&x ~,~
q_2(x,z_1,z_2)=z_2\cdot x +z_1^2
\end{eqnarray}
and a constant scale function $s(x,z_1,z_2)= 0.5$ in model (\ref{mod}). Note that $q_1$ corresponds to the null hypothesis, while
$q_2$ represents  an alternative.
The  results of the bootstrap test for the median
 are listed in Table \ref{tab:2D} for the sample size $n=50$ and we observe
in these examples  similar satisfactory properties  as in the one-dimensional setting.

\vspace{.5cm}

{\bf Acknowledgements.}
The authors thank Martina
Stein, who typed parts of this manuscript with considerable
technical expertise.
This work has been supported in part by the
Collaborative Research Center ``Statistical modeling of nonlinear
dynamic processes'' (SFB 823, Teilprojekt C1, C4) of the German Research Foundation
(DFG).

\bibliographystyle{apalike}
\bibliography{omission}

\begin{appendix}

\section{Appendix: Proofs} \label{sec6}
\def\theequation{A.\arabic{equation}}
\setcounter{equation}{0}

Throughout this section, introduce the abbreviation $\Theta_n := \Theta\cap \D_n$ with $\D_n := \{x: [x-h_n,x+h_n] \subset R_X \}$.

\begin{lemma}\label{lem1} If assumptions \ref{as:k1}-\ref{as:b2} and \ref{as:f2}-\ref{as:f6} are satisfied, then
\[
\hat q_\tau(x) = q_\tau(x) - \frac{1}{f_{\eps|X}(0|x)} \int_{-1}^1 \kappa(v) \Delta_S(q_{\tau+vb_n}(x)|x) dv + o_P(n^{-1/2}) =: \hat q_{\tau,L}(x) + o_P(n^{-1/2})
\]
uniformly in $x \in \D_n$ where $\Delta_{S}(x,y)$ is defined in Lemma \ref{lem:propfdach} and has the property
\[
\sup_{v \in [-1,1], x \in \D_n} |\Delta_S(q_{\tau+vb_n}(x)|x)| = O_P\Big(d_n^s + \Big(\frac{\log n}{nh_n^d} \Big)^{1/2}\Big).
\]
Moreover, $\hat q_{\tau,L}(x)$ is, with probability tending to one, $d+1$ times continuously differentiable with derivatives bounded uniformly on $\D_n$.

\end{lemma}

\noindent{\bf Proof.}
Apply part (a) of Lemma \ref{lem:genlin} to $F_{Y|X}(\cdot|x)$ and part (c) of the same Lemma with $F_1(\cdot|x) = F_{Y|X}(\cdot|x), F_2(\cdot|x) = \hat F_{Y|X}(\cdot|x;p)$. Combined the results with Lemma \ref{lem:propfdach} yields the assertion.
\hfill $\Box$

\begin{lemma}\label{lem2}
If assumptions \ref{as:k1} - \ref{as:b2}, \ref{as:f2} - \ref{as:f7}, \ref{as:c1} and \ref{as:c2} are satisfied, then
\begin{eqnarray*}
&&\int f_{\eps|X}(0\mid s)(\hat q_\tau(s)-q_\tau(s))I\{s\in\Theta_n\}f_{X}(s)F_{Z|X,\eps}(z|s,0)\,ds\\
&=&  -\frac{1}{n}\sum_{i=1}^n\Big(I\{\eps_i\leq 0\}-\tau\Big)I\{X_i\in\Theta_n\}F_{Z|X,\eps}(z|X_i,0)+o_P(\frac{1}{\sqrt{n}})
\end{eqnarray*}
uniformly with respect to $\Theta\in\Xi, z \in R_Z$.
\end{lemma}

\noindent{\bf Proof.}
>From Lemma \ref{lem1} we obtain the representation
\bean
&& - \int f_{\eps|X}(0|s)(\hat q_\tau(s) - q_\tau(s))I\{s\in\Theta_n\} f_X(s)F_{Z|X,\eps}(z|s,0)ds
\\
&=& \int_{-1}^1 \kappa(v) \int \Delta_S(q_{\tau+vb_n}(s)|s) I\{s\in\Theta_n\} f_X(s)F_{Z|X,\eps}(z|s,0)dsdv + o_P(n^{-1/2})
\\
&=& \int_{-1}^1 \kappa(v) \int \frac{1}{nh_n^d}\sum_i \MM(s)
\Big(\Omega \Big(\frac{q_{\tau+vb_n}(s)-Y_i}{d_n}\Big) - F_{Y|X}(q_{\tau+vb_n}(s)|X_i) \Big)
\\
&& \quad \quad \times \Big(\KK_{h_n,0}(s-X_i),...,\KK_{h_n,\kb_{N_p,p}}(s-X_i) \Big)^t
\\
&& \quad \quad \times I\{s\in\Theta_n\}(I_1(X_i;\Theta_n,h_n)+I_2(X_i;\Theta_n,h_n)) f_X(s)F_{Z|X,\eps}(z|s,0)dsdv
~+ o_P(n^{-1/2}),
%\\
%&=& \int_{-1}^1 \kappa(v) \int_{[-1,1]^d} \frac{1}{n}\sum_i \MM(X_i + h_ns)
%\Big(\Omega \Big(\frac{q_{\tau+vb_n}(X_i+sh_n)-Y_i}{d_n}\Big) - F_{Y|X}(q_{\tau+vb_n}(X_i+sh_n)|X_i) \Big)\times
%\\
%&& \quad \quad  \times \Big(K_{1,0}(s),...,K_{1,k_{N_p,p}}(s) \Big)^t I_1(X_i,s,x)f_X(X_i+sh_n)F_{Z|X,\eps}(z|X_i+sh_n)dsdv
%\\
%&& \int_{-1}^1 \kappa(v)  \frac{1}{n}\sum_i \int\MM(X_i + h_ns)
%\Big(\Omega \Big(\frac{q_{\tau+vb_n}(X_i+sh_n)-Y_i}{d_n}\Big) - F_{Y|X}(q_{\tau+vb_n}(X_i+sh_n)|X_i) \Big)\times
%\\
%&& \quad \quad  \times \Big(K_{1,0}(s),...,K_{1,k_{N_p,p}}(s) \Big)^t I_2(X_i,s,x)f_X(X_i+sh_n)F_{Z|X,\eps}(z|X_i+sh_n)dsdv
%\\
%&& + o_P(n^{-1/2}),
\eean
where
\[
\MM(s) := e_1^t\Big(\sum_{j=0}^{n_f} (-1)^j\Big(\frac{\M(K)^{-1}}{f_X(x)} \sum_{1\leq |\mb| < n_f} h_n^{|\mb|}f_X^{(\mb)}(x)M_\mb \Big)^j \frac{\M(K)^{-1}}{f_X(x)}\Big)
\]
and
\bean
I_1(X;\Theta_n,h_n) &:=& I\{ \otimes_{j=1}^d[X(j)-h_n,X(j)+h_n] \subset \Theta_n\},%I\{c+h_n\leq X_i\leq x-h_n\},
\\
I_2(X;\Theta_n,h_n) &:=&  I\{\exists j: [X(j)-h_n,X(j)+h_n]\not\subset \Theta_n, \otimes_{j=1}^d[X(j)-h_n,X(j)+h_n] \cap \Theta_n \neq \emptyset\}.
\eean
We will now proceed to show that the first part in the above decomposition [i.e. the part containing $I_1$] determines the asymptotic expansion and establish at the end of the proof that the part corresponding to $I_2$ is asymptotically negligible. First, note that
\bean
&&\int \frac{1}{nh_n^d}\sum_i \MM(s)
\Big(\Omega \Big(\frac{q_{\tau+vb_n}(s)-Y_i}{d_n}\Big) - F_{Y|X}(q_{\tau+vb_n}(s)|X_i) \Big)
\\
&& \quad \quad \times \Big(\KK_{h_n,0}(s-X_i),...,\KK_{h_n,\kb_{N_p,p}}(s-X_i) \Big)^t
I\{s \in \Theta_n\}I_1(X_i;\Theta_n,h_n)f_X(s)F_{Z|X,\eps}(z|s,0)ds
\\
&=& \int_{[-1,1]^d} \frac{1}{n}\sum_i \MM(X_i + h_ns)
\Big(\Omega \Big(\frac{q_{\tau+vb_n}(X_i+sh_n)-Y_i}{d_n}\Big) - F_{Y|X}(q_{\tau+vb_n}(X_i+sh_n)|X_i) \Big)
\\
&& \quad \quad  \times \Big(\KK_{1,0}(s),...,\KK_{1,\kb_{N_p,p}}(s) \Big)^t I_1(X_i;\Theta_n,h_n)f_X(X_i+sh_n)F_{Z|X,\eps}(z|X_i+sh_n,0)ds.
\eean
Observe that every entry of $\MM$ is by assumption continuously differentiable with respect to $s$ and the derivative is uniformly bounded. The class of functions defined by
\[
\Big\{(x,y) \mapsto \Omega \Big(\frac{q_{\zeta}(x+a)-y}{d_n}\Big)\Big||a(j)|\leq 1, j=1,...,d, |\zeta -\tau|\leq \alpha \Big\}
\]
where $\alpha$ is a small positive number has covering numbers that satisfy the assumptions of part 1 of Lemma \ref{lem:base} in Appendix \ref{append_b}. This follows from   Lemma \ref{lem:entrnum} together with the fact that under the assumptions \ref{as:f2}, \ref{as:f6} the mapping $(\zeta,a) \mapsto q_{\zeta}(x+a)$ satisfies
\[
\sup_x |q_{\zeta_1}(x+a_1)-q_{\zeta_2}(x+a_2)| \leq C(|\zeta_1-\zeta_2| + \|a_1-a_2\|_\infty)
\]
for some finite constant $C$ (this inequality is a consequence  of the implicit function theorem). Moreover, it follows from the smoothness assumptions on $F_{Y|X}$ and the properties of
$\Omega$ that
\[
\sup_{|s|\leq 1, |v|\leq 1} \Big| \E\Big[\Omega \Big(\frac{q_{\tau+vb_n}(X_i+sh_n)-Y_i}{d_n}\Big) - F_{Y|X}(q_{\tau+vb_n}(X_i+sh_n)|X_i) \Big| X_i \Big] \Big| \leq R_n \quad a.s.,
\]
where $R_n$ is a nonrandom quantity of order $o(1/\sqrt n)$. Thus the smoothness properties of $F_{Z|X,\eps}, F_{Y|X}$ and $(\zeta,x) \mapsto q_\zeta(x)$ imply that by Lemma \ref{lem:entrnum} and Lemma \ref{lem:base} in Appendix \ref{append_b} we have
\bean
&& \frac{1}{n}\sum_i \MM(X_i + h_ns)
\Big(\Omega \Big(\frac{q_{\tau+vb_n}(X_i+sh_n)-Y_i}{d_n}\Big) - F_{Y|X}(q_{\tau+vb_n}(X_i+sh_n)|X_i) \Big)
\\
&& \quad\quad \times   (\KK_{1,0}(s),...,\KK_{1,\kb_{N_p,p}}(s)  )^t I_1(X_i;\Theta_n,h_n) f_X(X_i+sh_n)F_{Z|X,\eps}(z|X_i+sh_n,0)
\\
&=& \frac{1}{n}\sum_i \MM(X_i)\Big(\KK_{1,0}(s),...,\KK_{1,\kb_{N_p,p}}(s) \Big)^t I\{X_i\in\Theta_n\} f_X(X_i)F_{Z|X,\eps}(z|X_i,0)
\\
&& \quad\quad\quad\quad \times \Big(\Omega \Big(\frac{q_{\tau+vb_n}(X_i+sh_n)-Y_i}{d_n}\Big) - F_{Y|X}(q_{\tau+vb_n}(X_i+sh_n)|X_i) \Big) + o_P(n^{-1/2})
\eean
uniformly with respect to  $|v|\leq 1, s \in [-1,1]^d, \Theta \in \Xi$ and $z\in R_Z$. Finally, noting that
\[
\Omega \Big(\frac{q_{\tau+vb_n}(X_i+sh_n)-Y_i}{d_n}\Big) = \Omega \Big(\frac{q_{\tau+vb_n}(X_i+sh_n)-q_\tau(X_i)-\eps_i}{d_n}\Big)
\]
yields
\[
\sup_{v,s,i}\Big|\Omega \Big(\frac{q_{\tau+vb_n}(X_i+sh_n)-Y_i}{d_n}\Big) - I\{\eps_i \leq 0\}\Big| \leq \|\Omega\|_\infty I\{|\eps_i|\leq R_n\} \quad a.s.,
\]
where $R_n = O(h_n+b_n+d_n)$ is a non-random quantity. This, together with an application of Lemma \ref{lem:base}, shows that
\bean
&&\frac{1}{n}\sum_i \MM(X_i)(\KK_{1,0}(s),...,\KK_{1,\kb_{N_p,p}}(s) )^t I\{X_i\in\Theta_n\} f_X(X_i)F_{Z|X,\eps}(z|X_i,0)
\\
&& \quad\quad\quad\quad \times \Big(\Omega \Big(\frac{q_{\tau+vb_n}(X_i+sh_n)-Y_i}{d_n}\Big) - F_{Y|X}(q_{\tau+vb_n}(X_i+sh_n)|X_i) \Big)
\\
&=& \frac{1}{n}\sum_i \MM(X_i) (I\{\eps_i \leq 0\} - F_{\eps|X}(0|X_i) )(\KK_{1,0}(s),...,\KK_{1,\kb_{N_p,p}}(s) )^t
\\
&& \quad\quad\quad\quad \times I\{X_i\in\Theta_n\} f_X(X_i)F_{Z|X,\eps}(z|X_i,0) + o_P(n^{-1/2}).
\eean
In particular, noting that $F_{\eps|X}(0|X_i) = \tau$, the above result implies
\bean
&&\int f_{\eps|X}(0|s)(\hat q_\tau(s) - q_\tau(s))I\{s \in \Theta_n\} f_X(s)F_{Z|X,\eps}(z|s,0)ds
\\
&=& \frac{1}{n}\sum_i \MM(X_i) (I\{\eps_i \leq 0\} - \tau ) (\mu_0(K),...,\mu_{\kb_{N_p,p}}(K) )^t I\{X_i\in\Theta_n\} f_X(X_i)F_{Z|X,\eps}(z|X_i,0)
\\
&&+ o_P(n^{-1/2}),
\eean
where $\mu_\kb(K) := \int_{\R^d} \KK_{1,\kb}(u) du$. Now from the definition of $\MM$ it is easy to see that
\[
\MM(x) = e_1^t( M_0(x)^{-1} + h_nR_M(x)) = e_1^t\Bigl(\frac{\M(K)^{-1}}{f_X(x)} + h_nR_M(x)\Bigr)
\]
where $R_M$ denotes a vector whose entries are uniformly bounded and Lipschitz-continuous with respect to $x$. Thus applying Lemma \ref{lem:base} we obtain
\bean
&&\frac{1}{n}\sum_i \MM(X_i)  (I\{\eps_i \leq 0\} - \tau  )  (\mu_0(K),...,\mu_{\kb_{N_p,p}}(K)  )^t  I\{X_i\in\Theta_n\} f_X(X_i)F_{Z|X,\eps}(z|X_i,0)
\\
&=& \frac{1}{n}\sum_{i=1}^n  (I\{\eps_i \leq 0\} - \tau  )I\{X_i\in\Theta_n\}F_{Z|X,\eps}(z|X_i,0)
+ o_P(n^{-1/2}),
\eean
which completes the first part of the proof.\\
It remains to show that
\bean
&&\frac{1}{n}\sum_i I_2(X_i;\Theta_n,h_n) \int_{-1}^1 \kappa(v) \int \frac{1}{h_n^d} \MM(s)
\Big(\Omega \Big(\frac{q_{\tau+vb_n}(s)-Y_i}{d_n}\Big) - F_{Y|X}(q_{\tau+vb_n}(s)|X_i) \Big)
\\
&& \quad \quad \times  (\KK_{h_n,0}(s-X_i),...,\KK_{h_n,\kb_{N_p,p}}(s-X_i)  )^t I\{s\in\Theta_n\} f_X(s)F_{Z|X,\eps}(z|s,0)dsdv
=  o_P(n^{-1/2})
\eean
uniformly with respect to $\Theta \in \Xi, z \in R_Z$. To this end, consider the ($n$-dependent) class of functions $\ef_n$ with elements
\begin{eqnarray*}
f_{z,\Theta_n,h_n,b_n}(x,y) &=& \int_{-1}^1 \kappa(v) \int \frac{1}{h_n^d} \MM(s)
\Big(\Omega \Big(\frac{q_{\tau+vb_n}(s)-y}{d_n}\Big) - F_{Y|X}(q_{\tau+vb_n}(s)|x) \Big)
\\
&  \times&   (\KK_{h_n,0}(s-x),...,\KK_{h_n,\kb_{N_p,p}}(s-x)  )^t I\{s\in\Theta_n\} f_X(s)F_{Z|X,\eps}(z|s,0)dsdv
\end{eqnarray*}
indexed by $z \in \Ze, \Theta \in \Xi$ contains uniformly bounded elements (the bound is also uniform with respect to $n$). Moreover, there exists a finite positive constant $C$ such that
\beq \label{eq:bbn}
N_{[\, ]}(\ef_n,\eps,L^2(P_X)) \leq \Big(N_{[\, ]}(\ef_{n,1},\eps/C,L^2(P_X))N_{[\, ]}(\ef_{n,2},\eps/C,L^2(P_X))\Big)^2,
\eeq
where $\ef_{n,1} := \{s \mapsto I\{s \in \Theta_n\}|\Theta \in \Xi\}$ and $\ef_{n,2} := \{s \mapsto F_{Z|X,\eps}(z|s,\eps)|z \in \Ze\}$.
To see that this holds, observe the decomposition
\bean
&&f_{z,\Theta_n,h_n,b_n}(x,y) = f^{(1)}_{z,\Theta_n,h_n,b_n}(x,y) + f^{(2)}_{z,\Theta_n,h_n,b_n}(x,y)
\\
&&:=\frac{1}{h_n^d}\sum_{j=1}^2\int\int \kappa(v)I\{\|x-s\|_\infty\leq h_n\}f_X(s)g_{j,n}(x,y,s,v)I\{s\in\Theta_n\}F_{Z|X,\eps}(z|s,0)dsdv
\eean
where $g_{1,n}$ and $g_{2,n}$ denote non-positive and non-negative, uniformly bounded functions, respectively. Moreover, $g_{j,n}$ do not depend on $\Theta_n$ or $z$. Obviously, it suffices to bound the bracketing number of $\F_{j,n} := \{(x,y) \mapsto f^{(j)}_{z,\Theta_n,h_n,b_n}(x,y)\}$ for $j=1,2$ separately. If we denote by $\{[b_{L,j},b_{U,j}]\}$ a collection of $\eps-$brackets (with respect to $L^2(P_X)$) for $\{s \mapsto I\{s\in\Theta_n\} F_{Z|X,\eps}(z|s,0)\}$. Then a collection of $\eps/C$ brackets for $\ef_{n,2}$ (with respect to $L^2(P_{X,Y})$) is given by
\[
B_{K,j}(x,y) := \frac{1}{h_n^d}\int\int \kappa(v)I\{\|x-s\|_\infty\leq h_n\}f_X(s)g_{2,n}(x,y,s,v)b_{K,j}(s)dsdv, \quad K=U,L.
\]
To see this, observe that
\bean
&&\E[(B_{L,j}(X_1,Y_1) - B_{U,j}(X_1,Y_1))^2]
\\
&\leq& \int \int\int g_{2,n}^2(x,y,s,v)\frac{1}{h_n^d}\kappa(v)I\{\|x-s\|_\infty\leq h_n\}f_X(s)\kappa(v) dsdv
\\
&& \quad\times \int\int \kappa(v)\frac{1}{h_n^d}I\{\|x-s\|_\infty\leq h_n\}f_X(s)(b_{U,j}(s) - b_{L,j}(s))^2 dsdv f_{X,Y}(x,y)dxdy
\\
&\leq& C_1 \int f_X(s)(b_{U,j}(s) - b_{L,j}(s))^2 \int \frac{1}{h_n^d}I\{\|x-s\|_\infty\leq h_n\} f_X(x)dxds
\eean
for some finite constant $C_1$. A bound for $\ef_{n,2}$ can be derived by similar arguments. Thus (\ref{eq:bbn}) is established.
%is H\"{o}lder continuous in the parameter $z$ with constants that, after possibly enlarging them, can be made independent of $n$ and that the elements of $\ef_n$ are uniformly bounded (also with respect to $n$). Thus we have $N_{[\, ]}(\ef_n,\eps,L^2(P_X))\leq C\eps^{-a}$ for some constants $C,a$ independent of $n$, see part 6 of Lemma \ref{lem:entrnum}.
Combining the bound in (\ref{eq:bbn}) with the assumptions \ref{as:c1} and \ref{as:c2}, the estimate
$
\sup_{z,\Theta}|\E[f_{z,\Theta_n,h_n,b_n}(X_1,Y_1)]| = o(n^{-1/2}),
$
and the results from Lemma \ref{lem:entrnum} and Lemma \ref{lem:base} yields the assertion after noting that by assumption $\sup_{\Theta \in \Xi} \E I_2(X_i;\Theta_n,h_n) = o(1)$. \hfill $\Box$

%\begin{lemma}\label{lem3}
%Under the assumptions of Theorem \ref{theo1} it holds that
%\begin{eqnarray*}
%\int (\hat q_\tau(s)-q_\tau(s))^2 f_X(s)\,ds &=&o_p(\frac{1}{\sqrt{n}})
%\end{eqnarray*}
%where $f_{X}$ denotes the density of $X$.
%\end{lemma}

%{\bf Proof.}

%\dotfill
%\hfill $\Box$

\begin{lemma}\label{lem4}
Under the assumptions of Theorem \ref{theo1} it holds that
\begin{eqnarray*}
T_n(\Theta_n,z) &=& \frac{1}{n}\sum_{i=1}^n (I\{\eps_i\leq 0\}-\tau )I\{X_i\in\Theta_n\}I\{Z_i\leq z\} +o_p(n^{-1/2}) \\
&&{}+\int  (F_{\eps|X,Z}(\hat q_{\tau,L}(s)-q_\tau(s)| s,t)-F_{\eps|X,Z}(0|s,t) )I\{s\in\Theta_n\}I\{t\leq z\}dF_{X,Z}(s,t),
\end{eqnarray*}
uniformly with respect to $\Theta\in\Xi, z\in R_Z $, where $F_{X,Z}$ denotes the joint distribution function of $X,Z$.
\end{lemma}

{\bf Proof.}
Note that
$
T_n(\Theta,z)=\frac{1}{n}\sum_{i=1}^n (I\{\hat\eps_i\leq 0\}-\tau )I\{X_i\in\Theta\}I\{Z_i\leq z\},
$
and that the assertion is equivalent to
\begin{eqnarray*}
&&\sup_{\Theta\in\Xi, z\in\Ze }\Bigg|\frac{1}{n}\sum_{i=1}^n (I\{\hat\eps_i\leq 0\}-I\{\eps_i\leq 0\} )I\{X_i\in\Theta_n\}I\{Z_i\leq z\}\\
&&\  - E\Big[ (I\{\hat\eps_L\leq 0\}-I\{\eps\leq 0\} )I\{X\in\Theta_n\}I\{Z\leq z\}\;\Big|\; (Y_i,X_i,Z_i)_{i=1,\dots,n}\Big]\Bigg|
 = o_p(\frac{1}{\sqrt{n}}).
\end{eqnarray*}

Here we define $\hat\eps_i=Y_i-\hat q_\tau(X_i)$, $\hat\eps_L=Y-\hat q_{\tau,L}(X)$, where we assume that
the sample $(Y_i,X_i,Z_i)$, $i=1,\dots,n$, (used to build $\hat q_{\tau,L}$) is independent from the generic variable $(Y,X,Z)$.
The proof now proceeds in two steps. First, note that by Lemma \ref{lem1} we have $\hat q_\tau - \hat q_{\tau,L} = o_P(n^{-1/2})$ uniformly on $\D_n$ and thus there exists a deterministic sequence $\gamma_n = o(n^{-1/2})$ with
\beq \label{eq:ql}
P(\sup_{x \in \D_n} |\hat q_\tau(x) - \hat q_{\tau,L}(x)| \leq \gamma_n) \rightarrow 1.
\eeq
Now on the set $\{|\hat q_\tau(x) - \hat q_{\tau,L}(x)| \leq \gamma_n\}$, the probability of which tends to one, we have
\[
\sup_{\Theta\in\Xi, z\in\Ze }\Bigg|\frac{1}{n}\sum_{i=1}^n (I\{\hat\eps_i\leq 0\}-I\{\hat\eps_{i,L}\leq 0\} )I\{X_i\in\Theta_n\}I\{Z_i\leq z\}\Bigg|
\leq \frac{1}{n}\sum_{i=1}^n I\{|\hat\eps_{i,L} |\leq \gamma_n\}I\{X_i \in \D_n\}
\]
Next, note that $I\{|\hat\eps_{i,L}| \leq \gamma_n\}=I\{|\eps_i - g(X_i)| \leq \gamma_n\}$ for $g=\hat q_{\tau,L}-q_\tau$.
Now the assertion follows since the ($n$-dependent) class of functions
\[
\Big\{(\epsilon,\xi)\mapsto I\{|\epsilon - g(\xi)| \leq \gamma_n \}I\{\xi\in\D_n\}\;\Big|\; g\in C_1^{d+1}(R_X)\Big\}
\]
satisfies the assumptions of part 1 of Lemma \ref{lem:base} whenever $n$ is sufficiently large, see the proof of Lemma A.3 in \cite{neukei2010} for a similar reasoning, and $\hat q_{\tau,L}-q_\tau\in C_1^{d+1}(\D_n)$ with probability converging to one by Lemma \ref{lem1}.
Here $C_1^{d+1}(\D_n)$ is the class of $d+1$ times differentiable functions $g$ defined on $\D_n$. Further, note that
\[
\sup_{g \in C_1^{d+1}(\D_n)} \E\Big[I\{|\eps_i - g(X_i)| \leq \gamma_n\}I\{X_i \in \D_n\} \Big] = o(n^{-1/2})
\]
This, together with (\ref{eq:ql}), and an application of Lemma \ref{lem:base}, shows that
\[
\sup_{\Theta\in\Xi, z\in\Ze }\Bigg|\frac{1}{n}\sum_{i=1}^n (I\{\hat\eps_i\leq 0\}-I\{\hat\eps_{i,L}\leq 0\} )I\{X_i\in\Theta_n\}I\{Z_i\leq z\}\Bigg|
= o_P(n^{-1/2}).
\]
Similar arguments applied to the (n-dependent) class functions
$$
\Big\{(\epsilon ,\xi ,\zeta )\mapsto  (I\{\epsilon\leq g(\xi)\}-I\{\epsilon\leq 0\} )I\{\xi\in\Theta_n\}I\{\zeta\leq z\}\;\Big|\; g\in C_1^{d+1}(R_X),\Theta\in \Xi,z\in \Ze\Big\}
$$
yield
\begin{eqnarray*}
&&\sup_{\Theta\in\Xi, z\in\Ze }\Bigg|\frac{1}{n}\sum_{i=1}^n (I\{\hat\eps_{i,L}\leq 0\}-I\{\eps_i\leq 0\} )I\{X_i\in\Theta_n\}I\{Z_i\leq z\}\\
&&\  - E\Big[ (I\{\hat\eps_L\leq 0\}-I\{\eps\leq 0\} )I\{X\in\Theta_n\}I\{Z\leq z\}\;\Big|\; (Y_i,X_i,Z_i), i=1,\dots,n\Big]\Bigg|
 = o_p(\frac{1}{\sqrt{n}}).
\end{eqnarray*}
and thus the proof is complete. \hfill $\Box$\\
\\

\noindent {\bf Proof of Theorem \ref{theo1}.}
Starting from the stochastic expansion given in Lemma \ref{lem4} we obtain by Taylor's expansion
\begin{eqnarray*}
T_n(\Theta_n,z) &=& \frac{1}{n}\sum_{i=1}^n(I\{\eps_i\leq 0\}-\tau)I\{X_i\in\Theta_n\}I\{Z_i\leq z\}\\
&&{}+\int f_{\eps|X,Z}(0| s,t)(\hat q_\tau(s)-q_\tau(s))I\{s\in\Theta_n\}I\{t\leq z\}dF_{X,Z}(s,t)\\
&&{}+ \int f'_{\eps|X,Z}(\xi_{x,s,n}| s,t)(\hat q_\tau(s)-q_\tau(s))^2I\{s\in\Theta_n\}I\{t\leq z\}dF_{X,Z}(s,t) +o_p(\frac{1}{\sqrt{n}})
\end{eqnarray*}
for some $\xi_{x,s,n}$ between $0$ and $\hat q_\tau(s)-q_\tau(s)$ where the last line is of order $o_p(n^{-1/2})$ due to Lemma \ref{lem1} and the assumptions $\sup_{x\in \D,y\in\R,z \in R_Z} |f'_{\eps|X,Z}(y| x,z)|<\infty$, $d_n^{2s} + \log n/nh_n^d = o(n^{-1/2})$.
Note that
\begin{eqnarray*}
&&\int f_{\eps|X,Z}(0| s,t)(\hat q_\tau(s)-q_\tau(s))I\{s\in\Theta_n\}I\{t\leq z\}dF_{X,Z}(s,t)\\
&=& \int F_{Z|X,\eps}(z|s,0)f_{\eps|X}(0|s)f_X(s)(\hat q_\tau(s)-q_\tau(s))I\{s\in\Theta_n\}ds.
\end{eqnarray*}
By Lemma \ref{lem2} we thus have
\begin{eqnarray*}
T_n(\Theta_n,z)  =  \frac{1}{n}\sum_{i=1}^n (I\{\eps_i\leq 0\}-\tau )I\{X_i\in\Theta_n\}\Big(I\{Z_i\leq z\}-F_{Z|X,\eps}(z|X_i,0)\Big)
  +o_p(\frac{1}{\sqrt{n}}).
\end{eqnarray*}
This completes the proof. \hfill $\Box$\\
\\
{\bf Proof of Corollary \ref{cor1} and \ref{cor2}.}
Define the sequence of $n$-dependent classes of functions
$$
\ef_n := \Big\{ (e,\xi,\zeta)\mapsto e I\{\xi\in\Theta\cap \D_n\}(I\{\zeta\leq z\}-F_{Z|X,\eps}(z|\xi,0))\;\Big|\; \Theta \in \Xi, z\in R_Z \Big\}
$$
and note that it is indexed by the totally bounded metric space $(\Xi\times R_Z,\rho)$ with metric
\[
\rho((\Theta_1,y),(\Theta_2,z)) := (\E[(W_{\Theta_1,y}-W_{\Theta_2,z})^2])^{1/2}
\]
where $W_{\Theta,z} := (I\{\eps_1\leq 0\} - \tau)I\{X_1\in\Theta\}(I\{Z_1\leq z\}-F_{Z|X,\eps}(z|X_1,0))$.
Moreover, it satisfies the assumptions of part 2 of Lemma \ref{lem:base}. A simple calculation in combination with the assumption $\sup_{\Theta \in \Xi} P(X_i\in \Theta\backslash\Theta_n ) = o(1)$ shows that all the assumptions of Theorem 2.11.23 in \cite{vaarwell1996} are satisfied. In particular, the covariances $\Cov(W_{\Theta_n,y},W_{\Theta_n',z})$ converge to $k(\Theta,y,\Theta',z)$ given in Corollary \ref{cor1}. This implies that the process
\begin{eqnarray*}
&& \sqrt{n}\Big(T_n(\Theta_n,z) - \tilde T_n(\Theta_n,z)\Big)
\\
&=& \frac{1}{n}\sum_{i=1}^n \Big((I\{\eps_i\leq 0\}-\tau )I\{X_i\in\Theta_n\} (I\{Z_i\leq z\}-F_{Z|X,\eps}(z|X_i,0) ) - \tilde T_n(\Theta_n,z)\Big)
 +o_p(\frac{1}{\sqrt{n}}).
\end{eqnarray*}
where
$
\tilde T_n(\Theta_n,z) := \E\Big[(I\{\eps_i\leq 0\}-\tau )I\{X_i\in\Theta_n\} (I\{Z_i\leq z\}-F_{Z|X,\eps}(z|X_i,0) )\Big]
$
converges weakly to the centered Gaussian process $T(\Theta_n,z)$ described in Corollary \ref{cor1}. Thus Corollary \ref{cor1} and \ref{cor2} follow after a straightforward calculation of the expectation $\tilde T_n(\Theta_n,z)$. Now the proof is complete.
\hfill $\Box$

\section{Technical results} \label{append_b}
\def\theequation{B.\arabic{equation}}
\setcounter{equation}{0}

Before stating the main results of this section, we discuss some basic properties of the local polynomial estimator $\hat F_{Y|X}(y|x;p)$. To this end, we note that
\[
\XX^t\WW\YY =  (V_{n,0}(x,y),V_{n,\kb_{1,1}}(x,y),...,V_{n,\kb_{N_p,p}}(x,y)  )^t
\]
with
\[
V_{n,\kb}(x,y) := \frac{h_n^{|\kb|}}{nh_n^d}\sum_{i=1}^n \KK_{h,\kb}(x-X_i)\Omega \Big(\frac{y-Y_i}{d_n}\Big).
\]

%Also, set
%\[
%V_{n,\kb,U}(x,y) := \frac{1}{nh_n^d}\sum_{i=1}^n \KK_{h,\kb}(x-X_i)I\{Y_i \leq y\}
%\]
%and define $\hat F_{Y,U}$ with $V_{n,\kb}$ in the definition of $\YY$ replaced by $V_{n,\kb,U}$. Note that by definition, $V_{n,\kb,S}(x,\cdot)$ is a smoothed version of $V_{n,\kb,U}(x,\cdot)$ that is obtained by convolution with the function $\frac{1}{d_n}\omega(\cdot /d_n)$ where $\omega := \Omega'$. In particular, this implies
%\bean
%V_{n,\kb,S}(x,y) &=& \int V_{n,\kb,U}(x,y-d_nu)\kappa(u)du,
%\\
%\partial_x^\mb\partial_y^l V_{n,\kb,S}(x,y) &=& \frac{1}{d_n^{l+1}}\int \kappa^{(l)}(u/d_n)\partial_x^\mb V_{n,\kb,U}(x,y-d_nu)du.
%\eean

\begin{lemma}\ \label{lem:propfdach}
Under the assumptions \ref{as:k1}, \ref{as:k2}, \ref{as:b1}, \ref{as:f2}, \ref{as:f3} it holds that
\bean
&&\hat F_{Y|X}(y|x;p) - F_{Y|X}(y|x)
\\
&=&
e_1^t\Big(\sum_{j=0}^{n_f} \Big(-\frac{\M(K)^{-1}}{f_X(x)} \sum_{1\leq |\mb| < n_f} h_n^{|\mb|}f_X^{(\mb)}(x)M_\mb \Big)^j \frac{\M(K)^{-1}}{f_X(x)}\Big) (T_{n,0,S}(x,y),...,T_{n,\kb_{N_p,p},S}(x,y)  )^t
\\
&& + o_P(n^{-1/2})
\\
&=:& \Delta_{S}(y|x) + o_P(n^{-1/2}) = O_P(d_n^s + \Big(\frac{\log n}{nh_n^d} \Big)^{1/2})
\eean
uniformly with respect to $(x,y)\in \D_n \times\Y$, where $\Y$ is any bounded subset of $\R$ and $M_\kb$ denote some matrices with uniformly bounded entries that are independent of $x,n,y$ and
\[
T_{n,\kb,S}(x,y) := \frac{1}{nh_n^d}\sum_i \KK_{h_n,\kb}(x-X_i)\Big(\Omega \Big(\frac{y-Y_i}{d_n}\Big) - F_{Y|X}(y|X_i) \Big).
\]
Moreover, the quantity $\Delta_{S}(y|x)$ is, with probability tending to one, $d+1$ times continuously differentiable with respect to $x$ and $y$ and all its partial derivatives of corresponding orders are uniformly bounded on $\D_n\times\Y$.

\end{lemma}
\noindent
\textbf{Proof.}
At the end of the proof, we will establish the following two representations
\begin{equation} \label{eq:fl1}
\quad \hat F_{Y|X}(y|x;p) = F_{Y|X}(y|x) + e_1^t(\XX^t\WW\XX)^{-1}  ( h_n^0 T_{n,0,S}(x,y),..., h_n^p T_{n,\kb_{N_p,p},S}(x,y)  )^t
 + O_P(h_n^{p+1}),
\end{equation}
\beq
\label{eq:fl2}
\quad (\XX^t\WW\XX)^{-1} = \HH^{-1}\Big(\sum_{j=0}^{n_f} \Big(-\frac{\M(K)^{-1}}{f_X(x)} \sum_{1\leq |\lb| < n_f} h_n^{|\lb|}M_\lb f_X^{(\lb)}(x) \Big)^j \frac{\M(K)^{-1}}{f_X(x)} +  1_{N\times N} O_P(h_n^{n_f})\Big)\HH^{-1},
\eeq
where $M_0,...,M_{\kb_{N_{n_f},n_f}}$ denote some matrices that do not depend on $n,x$, $M_0 = \M(K)$ is invertible,  $\HH$ is a diagonal matrix with entries $1,h_n,..,h_n,h_n^2,...,h_n^2,...,h_n^p,...,h_n^p$ and the term $h_n^{|\kb|}$ appears $N_\kb$ times in this vector.
By definition we have
\bean
\partial_y^r \partial_x^\mb T_{n,\kb,S}(x,y) &=& \frac{1}{nh_n^{d+|\mb |}}\sum_i \KK_{h_n,\kb}^{(\mb )}(x-X_i)\Big(\frac{1}{d_n^{r}}\omega^{(r-1)}\Big(\frac{y-Y_i}{d_n}\Big) - F_{Y|X}^{(r)}(y|X_i)\Big),
\eean
and tedious but straightforward calculations including integration-by parts and substitutions yield the estimates
\bean
\sup_{(x,y)\in\D_n\times}\E[\partial_y^r \partial_x^\mb T_{n,\kb,S}(x,y)] &=& O(d_n^{s-r}),
\\
\sup_{(x,y)\in\D_n}\E[(\partial_y^r \partial_x^\mb T_{n,\kb,S}(x,y))^2] &=& O\Big(\frac{1}{n h_n^{d+2|\mb|}d_n^{0\vee (2r-1)}} \Big).
\eean
A combination of parts 1,2 and 6 of Lemma \ref{lem:entrnum} shows that, for every $n$, the class of functions
\[
\ef_n = \Big\{(u,v)\mapsto \KK_{h_n,\kb}^{(\mb )}(x-u)\Big(\frac{1}{d_n^{r}}\omega^{(r-1)}\Big(\frac{y-v}{d_n}\Big) - F_{Y|X}^{(r)}(y|u)\Big) \Big| x \in R_X, y \in \R \Big\}
\]
satisfies the assumptions of part 2 of Lemma \ref{lem:base} with constants not depending on $n$, which, together with the above estimates  gives
\beq \label{eq:fl3}
\sup_{(x,y)\in \D} |\partial_y^r \partial_x^\mb T_{n,\kb,S}(x,y)| = O_P\Big(\frac{\log n}{n h_n^{d+2|\mb|}d_n^{0\vee (2r-1)}} \Big)^{1/2} + O(d_n^{s-r}).\\
\eeq
\\
Combining (\ref{eq:fl1}), (\ref{eq:fl2}) and (\ref{eq:fl3}) yields
\bean
&&e_1^t(\XX^t\WW\XX)^{-1}  (h_n^0 T_{n,0,S}(x,y),...,h_n^p T_{n,\kb_{N_p,p},S}(x,y)  )^t
\\
&=& e_1^t\Big(\sum_{j=0}^{n_f} \Big(\frac{\M(K)^{-1}}{f_X(x)} \sum_{1\leq |\lb| < n_f} h_n^{|\lb|}M_\lb f_X^{(\lb)}(x) \Big)^j \frac{\M(K)^{-1}}{f_X(x)}\Big)
 ( T_{n,0,S}(x,y),...,T_{n,\kb_{N_p,p},S}(x,y)  )^t
  + o_P(n^{-1/2}),
\eean
and thus the proof of the first part of the Lemma is complete.\\
\\
For a proof of the differentiability results, note that the $d+1-$fold differentiability of the product of every entry of a scalar product between two vectors follows from the $d+1-$fold differentiability of every entry of both vectors. This establishes that $\Delta_{S}(y|x)$ is $d+1$ times continuously differentiable with respect to both components and that all partial derivatives are uniformly bounded. By the results in (\ref{eq:fl3}) the proof is thus complete once we establish (\ref{eq:fl1}) and (\ref{eq:fl2}).\\
\\
\textit{Proof of (\ref{eq:fl1})} A Taylor expansion of $F_{Y|X}(y|x)$ gives
\bean
&&\frac{1}{nh_n^d} \sum_i \KK_{h_n,\kb}(x-X_i)F_{Y|X}(y|x)
\\
&=& \frac{1}{nh_n^d} \sum_{0 \leq |\mb|\leq p}\frac{\partial_x^\mb F_{Y|X}(y|X_i)}{\mb!}h_n^{|\mb|} \sum_i \KK_{h_n,\kb+\mb}(x-X_i) + O_P(h_n^{|\mb|+p+1}).
\eean
This fact, combined with
\[
\frac{e_1^t (\XX^t\WW\XX)^{-1}}{nh_n^d}
\left(
\begin{array}{c}
h_n^{|\mb|} \sum_i K_{h_n,\mb}(x-X_i)\\
\vdots\\
h_n^{p+|\mb|} \sum_i K_{h_n,\kb_{N_p,p}+\mb}(x-X_i)
\end{array}
\right)
= I\{\mb=0\},
\]
yields the representation
\bean
F_{Y|X}(y|x)&=&\frac{e_1^t (\XX^t\WW\XX)^{-1}}{nh_n^d}
\left(
\begin{array}{c}
h_n^0\sum_i K_{h_n,0}(x-X_i)F_{Y|X}(y|x)\\
\vdots\\
h_n^p\sum_i K_{h_n,\kb_{N_p,p}}(x-X_i)F_{Y|X}(y|x)
\end{array}
\right)\\
&=&
\frac{e_1^t(\XX^t\WW\XX)^{-1}}{nh_n^d}
\left(
\begin{array}{c}
h_n^0\sum_i K_{h_n,0}(x-X_i)F_{Y|X}(y|X_i)\\
\vdots\\
h_n^p\sum_i K_{h_n,\kb_{N_p,p}}(x-X_i)F_{Y|X}(y|X_i)
\end{array}
\right)
+ O_P(h_n^{p+1})
\eean
once we note that $\frac{1}{nh_n^d}\sum_i |K_{h_n,\kb_{N_p,p}}(x-X_i)|=O_P(1)$ and $e_1^t(\XX^t\WW\XX)^{-1} = (O_P(1),...,O_P(h_n^{-p}))$ [see the last part of the proof]. Thus
\[
\hat F_{Y|X}(y|x) = F_{Y|X}(y|x) +  e_1^t(\XX^t\WW\XX)^{-1}  (h_n^0 T_{n,0,S}(x,y),...,h_n^p T_{n,\kb_{N_p,p},S}(x,y)  )^t + O_P(h_n^{p+1}).
\]
\\
\textit{Proof of (\ref{eq:fl2})} The elements of the matrix $\XX^t\WW\XX$ are of the form
\bean
(\XX^t\WW\XX)_{k,l} = \frac{1}{nh_n^d}\sum_i \KK_{h_n,0}(x-X_i)(x-X_i)^\mb = \frac{h_n^{|\mb|}}{nh_n^d}\sum_i \KK_{h_n,\mb}(x-X_i)
\eean
where $\mb = \mb_1 + \mb_2$ and $\mb_1,\mb_2$ denote the $k'$th and $l'$th entry in the tuple of vectors $(0,\kb_{1,1},...,\kb_{N_1,1}, \kb_{1,2}, ...,\kb_{N_p,p})$, respectively. In particular, $d+1+n_f$-fold continuous differentiability of $f_X$ implies that
\[
\frac{1}{nh_n^d}\sum_i \KK_{h_n,\kb}(x-X_i) = \sum_{|\lb| < n_f}\mu_{|\kb|+|\lb|}(\KK)h_n^{|\lb|}f_X^{(\lb)}(x)  + O_P(\Big(\frac{\log n}{nh_n^d} \Big)^{1/2} + h_n^{n_f}).
\]
Thus we obtain a representation of the form
\[
\XX^t\WW\XX = \HH\Big(\sum_{|\lb| < n_f} h_n^{|\lb|}M_\lb f_X^{(\lb)}(x) + 1_{N\times N}O_P(h_n^{n_f})\Big)\HH
\]
where $M_0,...,M_{\kb_{N_{n_f},n_f}}$ denote some matrices that do not depend on $n,x$, $M_0 = \M(K)$ is invertible and $\HH$ is a diagonal matrix with entries $1,h_n,..,h_n,h_n^2,...,h_n^2,...,h_n^{p},...,h_n^{p}$ where the term $h_n^{|\kb|}$ appears $N_{|\kb|}$ times in this vector [see the definition at the beginning of the section]. Thus for $h_n$ sufficiently small an application of the Neumann series yields (\ref{eq:fl2}) with probability tending to one.
\hfill $\Box$

\begin{lemma}\label{lem:entrnum} \ \textbf{Bounds on bracketing numbers}
\begin{enumerate}
\item \label{entrnum0} Define $\ef+\G := \{f+g|f\in\ef,g\in\G\}, \ef\G := \{fg|f\in\ef,g\in\G\}$. Then
\[
N_{[\,]}(\ef+\G, \eps, \rho)\leq N_{[\,]}(\ef, \eps/2, \rho)N_{[\,]}(\G, \eps/2, \rho)
\]

If additionally the classes $\ef,\G$ are uniformly bounded by the constant $C$, we have
\[
N_{[\,]}(\ef\G, \eps, \|.\|) \leq N_{[\,]}^2(\ef, \eps/4C, \|.\|)N_{[\,]}^2(\G, \eps/4C, \|.\|)
\]
for any seminorm $\|.\|$ with the additional property that $|f_2|\leq|f_2|$ implies $\|f_1\|\leq \|f_2\|$.

\item \label{entrnum2} Let  $\ef_n$ denote a class of functions $f_x$ indexed by the bounded interval $x\in[-A,A]$ which are  bounded by a given constant and have   support of the form $[x-h,x+h]$. If  $\sup_{f\in \mathcal F}|f(a)-f(b)|\leq C|a-b|h^{-k}$ for some universal constant $C$ we have $N_{[\,]}(\mathcal F_n, \eps, L^2(P_X)) \leq \gamma \eps^{-(2k+1)}$ provided that $P_X$ has a uniformly bounded density. Here $\gamma$ denotes a constant which does not depend on $n$.

\item \label{entrnum8} Consider the class of functions
\[
\ef_n := \Big\{ (x,y) \mapsto \Omega\Big(\frac{g(x)-y}{d_n} \Big) \Big| g \in \G \Big\},
\]
where $\Omega$ is Lipschitz-continuous and there exist constants $C_1,C_2$ such that $\Omega$ is constant on $(-\infty,C_1]$ and $[C_2,\infty)$. Assume additionally that the  distribution of $(X,Y)$ has a uniformly bounded density, then
\[
N_{[\,]}(\ef_n, \eps, L^2(P_{XY})) \leq C_5 N_{[\,]}(\G, C_6\eps^2, \|\cdot\|_\infty)
\]
for some constants $C_5,C_6$ independent of $n$.

\item \label{entrnum4} For any measure $P^{U,V}$ on the unit interval with uniformly bounded density $f$, the class of functions
\[
\ef :=  \{ u \mapsto I\{u\leq s\}  | s \in [0,1]  \} \cup  \{ u \mapsto I\{u < s\}  | s \in [0,1]  \}
\]
can be covered by $C\eps^{-2}$ brackets of $L^2(P)$ length $\eps$.

\item \label{entrnum7} For any measure $P$ on $\R\times\R^k$ with uniformly bounded conditional density $f_{V|U}$ the class of functions
\[
\G :=  \{(u,v) \mapsto I\{v \leq f(u)\}  | f \in \ef  \}
\]
satisfies $N_{[\,]}(\G, \eps, \|.\|_{P,2}) \leq N_{[\,]}(\ef, C\eps^2, \|.\|_\infty)$ for some constant $C$ independent of $\eps$.

\item \label{entrnum9} Assume that $f(x;a)$ is a function indexed by the parameter $a\in A$ such that $\sup_x \|f(s;x)- f(t;x)\|_\infty \leq C\|s-t\|^\theta$ for some $\theta>0$ and norm $\|\cdot\|$. Then the $\|\cdot\|_\infty$-bracketing numbers of the class of functions $\ef = \{u \mapsto f(u;a)|a\in A\}$ satisfy $N_{[\,]}(\ef,\eps,\|\|_\infty) \leq C_1 N(A,C_2\eps^{1/\theta},\|\cdot\|)$ for some finite constants $C_1, C_2$.

\end{enumerate}
\end{lemma}
\textbf{Proof.}\\ %\textbf{Stanislav: bitte die restlichen Teile auff\"{u}hren!!} \\
\textbf{Part \ref{entrnum0}}
The first assertion is obvious from the definition of bracketing numbers. For the second assertion, note that $\ef\G = (\ef+C)(\G+C) - C\ef - C\G + C^2.$ Moreover, all elements of the classes $\ef+C,\G+C$ are by construction non-negative and thus it also is possible to cover them with brackets consisting of non-negative functions and amounts equal to the brackets of $\ef,\G$, respectively. Finally, observe that if $0 \leq f_l \leq f \leq f_u$ and $0 \leq g_l \leq g \leq g_u$, we also have $f_lg_l \leq fg \leq f_ug_u$. Moreover $\|f_lg_l - f_ug_u\| \leq C\|f_u-f_l\| + C\|g_u-g_l\|$. Thus the class $(\ef+C)(\G+C)$ can be covered by at most $\leq N_{[]}(\ef, \eps, \|.\|)N_{[]}(\G, \eps, \|.\|)$ brackets of length $2C\eps$. Finding brackets for the classes $C\ef, C\G$ is trivial, and applying the first assertion of the Lemma completes the proof.\\
\textbf{Part \ref{entrnum2}}
Consider two cases. \\
% \begin{itemize} \item[
A) $\eps> 4h^{1/2}$:
Divide $[0,1]$ into $N := 2/\eps^2$ subintervals of length $2\alpha := \eps^2$ with centers $r\alpha$ for $r=1,...,N$ and call the intervals $I_1,...,I_N$. Note that two adjunct intervals overlap by $\alpha > 2h$. This construction ensures that every set of the form $[x-h,x+h]$ with $x\in[h,1-h]$ is completely contained in at least one of the intervals defined above. Then a collection of $N$ brackets of $L^2$-length $D\eps$ for some $D>0$ independent of $h$ is given by $(-CI\{u \in I_j\}, CI\{u \in I_j\})$. \\
%\item[
B)
$\eps \leq 4h^{1/2}$~:
Observe that by assumption any element $g$ of $\ef$ satisfies $|g(x)-g(y)|\leq C|x-y|h^{-k}$. Consider the points $t_i := i/(N+1), i=1,...,N$ with $N := 2^{2k+1}C/\eps^{2k+1}$. By construction, to every $x\in[h,1-h]$ there exists $i(x)$ with $|t_{i(x)}-x|\leq \eps^{2k+1}/(2^{2k+1}C)$. This implies
\[
|g(x) - g(t_{i(x)})| \leq C  \eps^{2k+1} h^{-k}/2^{2k+1}C \leq \eps/2
\]
Then $N$ $\|.\|_\infty-$brackets of length covering $\ef$ are given by $(g(t_i)-\eps/2,g(t_i)+\eps/2)$, $i=1,...,N$. From those one can easily construct $L^2(P_X)$-brackets.\\
%\end{itemize}
\textbf{Part \ref{entrnum8}}
Without loss of generality, assume that $\Omega$ equals one on $[1,\infty)$ and zero on $(-\infty,-1]$. Moreover, the assumptions on $\Omega$ imply the existence of finite constant $C_l,C_u$ such that $C_l \leq \Omega \leq C_u$. Distinguish two cases
% \begin{enumerate} \item[
A) $\eps \leq d_n$~:
Starting with $\eps^2$ supremum norm brackets for the class $\G$ and using the Lipschitz condition yields the desired brackets.
%\item[
B) $\eps > d_n$~:
Denote by $[g_{1,l},g_{1,u}],...,[g_{N(\eps),l},g_{N(\eps),u}]$ brackets for the class $\G$ of $\|\cdot\|_\infty$-size $\eps$. For a function $g\in\G$, denote the bracket that contains it by $[g_{j(g),l},g_{j(g),u}]$. Observe that
\[
\Omega\Big(\frac{g(x)-y}{d_n} \Big)
\left\{
\begin{array}{ccc}
= 0, & if & y>g_{j(g),u}(x)+d_n \\
= 1, & if & y<g_{j(g),l}(x)-d_n \\
\in [C_l,C_u] && else
\end{array}
\right.
\]
Thus brackets of the form
\bean
b_{l,j}(x) &:=& I\{y<g_{j,l}(x)-d_n\} + C_lI\{g_{j,l}(x) - d_n \leq y \leq g_{j,u}(x) + d_n\}
\\
b_{u,j}(x) &:=& I\{y<g_{j,l}(x)-d_n\} + C_uI\{g_{j,l}(x) - d_n \leq y \leq g_{j,u}(x) + d_n\}
\eean
contain every function in $\ef_n$. Moreover, the $L^2$-length of each such bracket is bounded by $(C_u-C_l)(2d_n+\eps)\sup f_{X,Y}(x,y) \leq C\eps$. This completes the proof.\\
%\end{enumerate}
\textbf{Part \ref{entrnum4}} Follows by standard arguments.
\\
\textbf{Part \ref{entrnum7}} Follows from $|I\{v \leq g_1(u)\} - I\{v \leq g_2(u)\}| \leq I\{|v-g_1(u)|\leq 2 \|g_1-g_2\|_\infty\}$.\\
\textbf{Part \ref{entrnum9}} Obvious \hfill $\Box$
%\textbf{Rest in Birke, Neumeyer, Volgushev 2012. - das paper gibt es noch nicht. Bitte den Beweis hier einf\"{u}gen!}

\begin{lemma}[Basic Lemma]\label{lem:base}\  Assume that the classes of functions $\ef_n$ consist of uniformly bounded functions (by a constant    not depending on $n$).
\begin{enumerate}
\item  If for some $a<2$  $N_{[\, ]}(\ef_n,\eps,L^2(P))\leq C\exp(-c\eps^{-a})$ for every $\eps\leq\delta_n$  with constants $C,c$ not depending on $n$, then we have
\[
\sqrt{n} \sup_{f\in \ef_n, \|f\|_{P,2} \leq \delta_n} \Big(\int f dP_n - \int f dP\Big) = o_P^*(1),
\]
where the $*$ denotes outer probability, see \cite{vaarwell1996} for a more detailed discussion.
\item If  $N_{[\, ]}(\ef_n,\eps,L^2(P))\leq C\eps^{-a}$ for every $\eps\leq \delta_n$, some $a>0$ and a constant $C$ not depending on $n$, then we have for any $\delta_n \sim n^{-b}$ with $b<1/2$
\[
\sqrt{n} \sup_{f\in \ef_n, \|f\|_{P,2} \leq \delta_n} \Big(\int f dP_n - \int f dP\Big) = O_P^*\Big( \delta_n |\log \delta_n| \Big).
\]
\end{enumerate}
\end{lemma}
\textbf{Proof.}
Start by observing that the uniform boundedness of elements of $\ef_n$ by $D$ implies that $F \equiv D$ is a measurable envelope function with $L_2$-norm $D$. The proof of the first part follows by arguments similar to those used for the proof of the second part and is therefore omitted. For the proof of the second part, note that for $\eta_n$ sufficiently small
\bean
a(\eta_n) &:=& \eta_n D/\sqrt{1 + \log N_{[ ]}(\eta_n D,\ef_n,L_2(P))} \geq D\eta_n /\sqrt{1 + \log C -a \log(D\eta_n)}
\\
&\geq& D\tilde C \eta_n/\sqrt{|\log \eta_n|}
\eean
for some finite constant $\tilde C$ depending only on $a,C,D$. Thus the bound in Theorem 2.14.2 in van der Vaart, Wellner (1996) yields for $\delta_n$ sufficiently small
\bean
\E\Big[\sup_{f\in\ef_n} \int f d\alpha_n\Big]^* &\leq& DJ_{[ ]}(\delta_n,\ef_n,L_2(P)) + \sqrt{n}\int F(u)I\{F(u)>\sqrt{n}a(\delta_n)\}P(du)
\\
&\leq& D C_1\int_0^{\delta_n} |\log \eps| d\eps + D\sqrt{n}I\Big\{D>\frac{D\tilde C \sqrt{n}\delta_n}{|\log\delta_n|} \Big\}
\\
&\leq& D C_2\delta_n|\log \delta_n| + D\sqrt{n}I\Big\{1>\frac{\tilde C \sqrt{n}\delta_n}{|\log\delta_n|} \Big\}.
\eean
where $\alpha_n := \sqrt{n}(P_n - P)$, $P_n$ denotes the empirical measure, and $C_1,C_2$ are some finite constants. Here, the second inequality follows by a straightforward calculation and the first inequality is due to the fact that for $\delta_n$ sufficiently small by definition
\bean
J_{[ ]}(\delta_n,\ef_n,L_2(P)) = \int_0^{\delta_n} \sqrt{1 + \log N_{[ ]}(\eps D,\ef_n,L_2(P))}d\eps \leq C_1 \int_0^{\delta_n} |\log \eps|d\eps.
\eean
Now under the assumption on $\delta_n$, the indicator in the last line will be zero for $n$ large enough and thus the proof is complete. \hfill $\Box$

\begin{lemma} \label{lem:genlin}
Assume that $\kappa$ is a symmetric, uniformly bounded density with support $[-1,1]$ and let $b_n = o(1)$. Introduce the notation $Q_{G,\kappa,\tau,b_n}(F):= G^{-1}(H_{G,\kappa,\tau,b_n}(F))$.\\
(a) If the function $F: [0,1]\rightarrow \R$ is strictly increasing and $F^{-1}$ is $k$ times continuously differentiable in a neighborhood of the point $\tau$, we have
\[
H_{id,\kappa, \tau,b_n}(F) = F^{-1}(\tau) + \sum_{i=1}^k \frac{b_n^i}{i!}(F^{-1})^{(i)}(\tau)\mu_{i+1}(\kappa) + R_n(\tau)
\]
with $|R_n(\tau)| \leq C_k(\kappa) b_n^k \sup_{|s-\tau|\leq b_n}|(F^{-1})^{(k)}(\tau)-(F^{-1})^{(k)}(s)|$, $\mu_i(\kappa) := \int u^i\kappa(u)du$ and a constant $C_k$ depending only on $k$ and $\kappa$. In particular, if  $F: \R\rightarrow [0,1]$ is strictly increasing and $F^{-1}$ is two times continuously differentiable in a neighborhood of $\tau$ and $G:[0,1]\rightarrow\R$ is two times continuously differentiable in a neighborhood of $F^{-1}(\tau)$ with $G'(F^{-1}(\tau))>0$, we have
\[
|F^{-1}(\tau) - Q_{G,\kappa,\tau,b_n}(F)| \leq R_{n,2}:= Cb_n^2\sup_{|s-G\circ F^{-1}(\tau)|\leq R_{n,1}}|(G^{-1})'(s)|\sup_{|s-\tau|\leq b_n }|(G\circ F^{-1})''(s)|
\]
for some constant $C$ that depends only on $\kappa$ where $R_{n,1} := Cb_n^2\sup_{|s-\tau|\leq b_n }|(G\circ F^{-1})''(s)|$.\\
\\
(b) Assume that $\kappa$ is additionally differentiable with Lipschitz-continuous derivative and that the functions $G,G^{-1}$ have derivatives that are uniformly bounded on any compact subset of $\R$ [the bound is allowed to depend on the interval].
Then for any increasing function $F$ with uniformly bounded first derivative we have $|H(F_1)-H(F_2)|\leq R_{n,3} + R_{n,4}$ and
\[
|Q_{G,\kappa,\tau,b_n}(F_1) - Q_{G,\kappa,\tau,b_n}(F_2)| \leq \sup_{u\in \mathcal{U}(H(F_1),H(F_2))}|(G^{-1})'(u)|(R_{n,3} + R_{n,4}),
\]
where the constant $C$   depends only on $\kappa$, $\mathcal{U}(a,b) := [a \wedge b, a \vee b]$, and
\[
R_{n,3} := \frac{Cc_n}{b_n}\|F_1-F_2\|_\infty \sup_{|v-\tau|\leq c_n }|(G\circ F^{-1})'(v)|, \quad R_{n,4} := R_{n,3} \frac{\|F_1-F\|_\infty + \|F_1-F_2\|_\infty}{b_n}
\]
with $c_n := b_n + 2\|F_1-F_2\|_\infty + \|F_1-F\|_\infty$.\\
\\
(c) If additionally to the assumptions made in (b), the function $F_1$ is two times continuously differentiable in a neighborhood of $F^{-1}(\tau)$ with $F_1'(F_1^{-1}(\tau )) > 0$  and $G$ is two times continuously differentiable in a neighborhood of $F_1^{-1}(\tau)$ with $G'(F^{-1}(\tau))>0$, we have
\bean
Q_{G,\kappa,\tau,b_n}(F_1) - Q_{G,\kappa,\tau,b_n}(F_2)
&=& -\frac{1}{F_1'(F_1^{-1}(\tau ))}\int_{-1}^1 \kappa(v)\Big(F_2(F_1^{-1}(\tau+vb_n)) - F_1(F_1^{-1}(\tau+vb_n)) \Big)dv
\\
&&+ R_n,
\eean
where
\bean
|R_n| &\leq& R_{n,5} + R_{n,6} + \frac{Cb_n \sup_{|s-\tau|\leq b_n}(G\circ F^{-1})''(s)\|F_1-F_2\|_\infty + R_{n,4}}{G'(F_1^{-1}(\tau))}
\eean
with a constant $C$ depending only on $\kappa$ and
\bean
R_{n,5} &:=& \frac{1}{2}\sup_{u\in \mathcal{U}(H(F_1),H(F_2))}|(G^{-1})''(u)|(H(F_1)-H(F_2))^2
\\
R_{n,6} &:=& \sup_{u\in \mathcal{U}(H(F_1),G(F_1^{-1})(\tau))}|(G^{-1})''(u)|\cdot |H(F_1) - G(F_1^{-1})(\tau)|\cdot |H(F_1)-H(F_2)|.
\eean
\end{lemma}

\noindent\textbf{Proof.}
The proof of the first part of (a) is essentially a Taylor expansion.    Details can be found in the proof of Lemma A.4 in \cite{volgushev2006}.
For a proof of the second part of (a), observe that by definition $H_{G,\kappa, \tau,b_n}(F) = H_{id,\kappa, \tau,b_n}(F\circ G^{-1})$. Together with the first part we obtain
\bean
|H_{id,\kappa, \tau,b_n}(F\circ G^{-1}) - G\circ F^{-1}(\tau)| &\leq&  Cb_n^2\sup_{|s-\tau|\leq b_n }|(G\circ F^{-1})''(s)| =: R_{n,1}
\eean
which yields
\bean
|G^{-1}(H_{G,\kappa, \tau,b_n}(F)) - F^{-1}(\tau)| &\leq& |(G^{-1})'(\xi)|\cdot|H_{id,\kappa, \tau,b_n}(F\circ G^{-1}) - G(F^{-1}(\tau))|
\\
&\leq& Cb_n^2\sup_{|s-G\circ F^{-1}(\tau)|\leq R_{n,1}}|(G^{-1})'(s)|\sup_{|s-\tau|\leq b_n }|(G\circ F^{-1})''(s)| =: R_{n,2}.
\eean
The proof of (a) is thus complete.\\
\\
>>>From now on, drop the index of $H$ for the sake of a simpler notation. For a proof of (b), observe the decomposition
\bean
H(F_1) - H(F_2) &=& - \frac{1}{b_n}\int_0^1  {\kappa}\Bigl(\frac{F_1(G^{-1}(u))-\tau}{b_n}\Bigr) (F_1(G^{-1}(u)) - F_2(G^{-1}(u))) du
\\
&& {}-  \frac{1}{b_n}\int_0^1 \Bigl [ {\kappa}\Bigl(\frac{\xi(u)-\tau}{b_n}\Bigr) - {\kappa}\Bigl (\frac{F_1(G^{-1}(u))-\tau}{b_n} \Bigr) \Bigr ]\\
&&{}\quad\times( F_1(G^{-1}(u)) - F_2(G^{-1}(u))) du
\eean
for some $|\xi(u) - F_2(G^{-1}(u))|\leq |F_1(G^{-1}(u))-F_2(G^{-1}(u))|$. This yields the bound
\bean
|H(F_1) - H(F_2)| &\leq& \frac{1}{b_n}\int_0^1  {\kappa} \Bigl ( \frac{F_1(G^{-1}(u))-\tau}{b_n} \Bigr) + \Big|  {\kappa} \Bigl ( \frac{\xi(u)-\tau}{b_n}\Bigr ) -  {\kappa} \Bigl( \frac {F_1(G^{-1}(u))-\tau}{b_n}\Bigr ) \Big|du
\\
&& \times \|F_1 - F_2\|_\infty
\eean
Next, observe that by assumption $\kappa$ is Lipschitz continuous and thus we have the inequality
\bean
&&\Bigl| {\kappa}\Bigl ( \frac{\xi(u)-\tau}{b_n} \Bigr) -  {\kappa} \Bigl ( \frac{F_1(G^{-1}(u))-\tau}{b_n}\Bigr ) \Bigr|
\\
&\leq& \frac{L|\xi(u)-F_1(G^{-1}(u))|}{b_n} \left(I{\{|F_1(G^{-1}(u))-\tau|\leq b_n\}}+I{\{|\xi(u)-\tau|\leq b_n\}}\right)
\\
&\leq& \frac{2L\|F_1 - F_2\|_\infty}{b_n}I{\{|F_1(G^{-1}(u))-\tau|\leq b_n + 2\|F_1-F_2\|_\infty\}}
\\
&\leq& \frac{2L\|F_1 - F_2\|_\infty}{b_n}I{\{|F(G^{-1}(u))-\tau|\leq b_n + 2\|F_1-F_2\|_\infty + \|F_1-F\|_\infty\}}.
\eean
Similarly
\bean
&&\Bigl|  {\kappa}\Bigl( \frac{F_1(G^{-1}(u))-\tau}{b_n} \Bigr) - {\kappa} \Bigl ( \frac {F(G^{-1}(u))-\tau}{b_n} \Bigr ) \Bigr|
\\
&\leq& \frac{2L\|F_1-F\|_\infty}{b_n}I{\{|F(G^{-1}(u))-\tau|\leq b_n + \|F_1-F\|_\infty\}},
\eean
and moreover
\[
\Bigl|  {\kappa}\Bigl( \frac {F(G^{-1}(u))-\tau}{b_n}\Bigr) \Bigr| \leq CI{\{|F(G^{-1}(u))-\tau|\leq b_n\}}.
\]
Define $c_n := b_n + 2\|F_1-F_2\|_\infty + \|F_1-F\|_\infty$. Note that the monotonicity of $F,G$ implies
\[
\{u:|F(G^{-1}(u))-\tau| \leq c_n\} \subseteq [G(F^{-1}(\tau - c_n)),G(F^{-1}(\tau + c_n))]
\]
and
\[
|G(F^{-1}(\tau + c_n)) - G(F^{-1}(\tau - c_n))| \leq 2c_n \sup_{|v-\tau|\leq c_n }|(G\circ F^{-1})'(v)|.
\]
In particular, this implies the estimate
\[
\int_0^1 I\{|F(G^{-1}(u))-\tau|\leq c_n\} du \leq 2c_n \sup_{|v-\tau|\leq c_n }|(G\circ F^{-1})'(v)|.
\]
Summarizing, we have obtained the bound $|H(F_1) - H(F_2)| \leq R_{n,3} + R_{n,4}$
where $C$ denotes some constant depending only on the kernel $\kappa$. Assertion (b) follows from this estimate and a Taylor expansion of $G^{-1}$.\\
\\
For a proof of assertion (c), note that after a substitution
\bean
&&\frac{1}{b_n}\int_0^1  {\kappa}\Bigl ( \frac {F_1(G^{-1}(u))-\tau}{b_n}\Bigr) (F_1(G^{-1}(u)) - F_2(G^{-1}(u))) du
\\
&=& \int_{-1}^1 (G\circ F_1^{-1})'(\tau + vb_n) \kappa(v)\Big(F_2(F_1^{-1}(\tau+vb_n)) - F_1(F_1^{-1}(\tau+vb_n)) \Big)dv
\\
&=& (G\circ F_1^{-1})'(\tau) \int_{-1}^1 \kappa(v)\Big(F_2(F_1^{-1}(\tau+vb_n)) - F_1(F_1^{-1}(\tau+vb_n)) \Big)dv + r_n
\eean
where
\[
|r_n| \leq C b_n \sup_{|s-\tau|\leq b_n}|(G\circ F_1^{-1})''(s)|\cdot \|F_1-F_2\|_\infty
\]
by a Taylor expansion of $(G\circ F_1^{-1})'$. A Taylor expansion of $G^{-1}$ yields
\bean
&&\Big\| G^{-1}(H(F_1)) - G^{-1}(H(F_2)) - (G^{-1})'(H(F_1))(H(F_1)-H(F_2))\Big\|
\\
&\leq& \frac{1}{2}\sup_{u\in \mathcal{U}(H(F_1),H(F_2))}|(G^{-1})''(u)|(H(F_1)-H(F_2))^2
\eean
where $\mathcal{U}(a,b) := [a \wedge b, a \vee b]$. A Taylor expansion yields
\[
\Big|(G^{-1})'(H(F_1)) - (G^{-1})'(G(F_1^{-1})(\tau)) \Big| \leq  \sup_{u\in \mathcal{U}(H(F_1),G(F_1^{-1})(\tau))}|(G^{-1})''(u)|\cdot|H(F_1) - G(F_1^{-1})(\tau)|
\]
and combining this with the results obtained so far we arrive at
\bean
&&\Big|Q(F_1) - Q(F_2) + \frac{1}{F_1'(F_1^{-1}(\tau))}\int_{-1}^1 \kappa(v)\Big(F_2(F_1^{-1}(\tau+vb_n)) - F_1(F_1^{-1}(\tau+vb_n)) \Big)dv \Big|
\\
&\leq& \Big\| G^{-1}(H(F_1)) - G^{-1}(H(F_2)) - (G^{-1})'(H(F_1))(H(F_1)-H(F_2))\Big\|
\\
&& + |H(F_1)-H(F_2)|\cdot|(G^{-1})'(H(F_1)) - (G^{-1})'(G\circ F_1^{-1}(\tau))|
\\
&& + \Big|\frac{H(F_1)-H(F_2)}{G'(F_1^{-1}(\tau))} +  \frac{1}{F_1'(F_1^{-1}(\tau))}\int_{-1}^1 \kappa(v)\Big(F_2(F_1^{-1}(\tau+vb_n)) - F_1(F_1^{-1}(\tau+vb_n)) \Big)dv \Big|
\\
&\leq& R_{n,5} + R_{n,6} + \frac{Cb_n \sup_{|s-\tau|\leq b_n}(G\circ F^{-1})''(s)\|F_1-F_2\|_\infty + R_{n,4}}{G'(F_1^{-1}(\tau))}.
\eean
This completes the proof.
\hfill $\Box$

\end{appendix}

\end{document}